\documentclass[10pt, onecolumn, a4paper, fleqn]{article}

\usepackage{STY/PACKAGES}
\usepackage{STY/DEFINITIONS}
\usepackage{STY/SETTINGS}
\usepackage{STY/TIKZ}
\usepackage{STY/TITLEPAGE}

\usepackage{mdframed} 

\begin{document}
\edef\myindent{\the\parindent}
\author{Kronberg, Vi \and Anthonissen, Martijn \and ten Thije Boonkkamp, Jan \and IJzerman, Wilbert}
\title{Three-Dimensional Freeform Reflector Design with a Microfacet Surface Roughness Model}

\thispagestyle{empty} 
\pagenumbering{gobble} 

\newgeometry{margin=19mm,top=30mm}

\begin{figure}[t]
	\centering
	\includegraphics[width=0.4\textwidth]{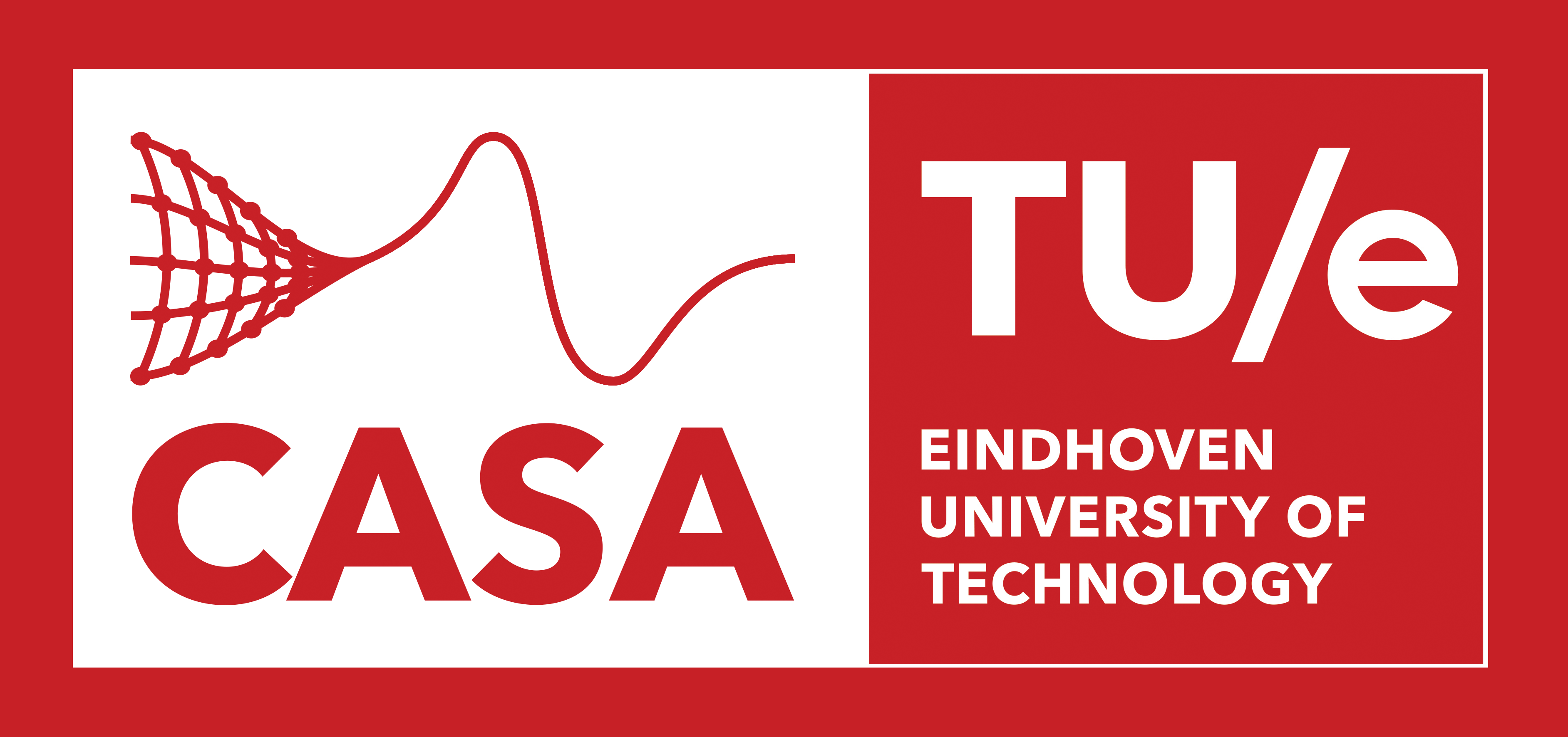}
\end{figure}

\vspace*{-0.7cm}
\begin{center}
	\LARGE \textbf{Three-Dimensional Freeform Reflector Design with a Microfacet Surface Roughness Model}\\[0.5cm]
\end{center}

\begin{center}
	\large
	{\itshape V\`{i}}~\textsc{Kronberg},\textsuperscript{1,*} {\itshape Martijn}~\textsc{Anthonissen},\textsuperscript{1}\\
	{\itshape Jan}~\textsc{ten Thije Boonkkamp},\textsuperscript{1} and {\itshape Wilbert}~\textsc{IJzerman}\textsuperscript{1,2}
\end{center}

\begin{changemargin}{1.5cm}{1.5cm}
\noindent
\textsuperscript{1}{\itshape Department of Mathematics and Computer Science, Eindhoven University of Technology},\\PO Box 513, 5600 MB Eindhoven, The Netherlands\\
\textsuperscript{2}{\itshape Signify Research}, High Tech Campus 7, 5656 AE Eindhoven, The Netherlands\\
\textsuperscript{*}{\color{blue}\href{mailto:v.c.e.kronberg@tue.nl}{\texttt{v.c.e.kronberg@tue.nl}}}\\
\url{https://www.win.tue.nl/~martijna/Optics/}\\
\end{changemargin}

\hrule
\begin{changemargin}{1.5cm}{1.5cm}
\textsc{\textbf{AMS:}} 78A05 $\cdot$ 78A45 $\cdot$ 78A46
\end{changemargin}
\hrule
\begin{changemargin}{1.5cm}{1.5cm}
	\textsc{\textbf{Abstract:}}
	This manuscript will unify inverse freeform reflector design and surface light scattering to design freeform reflectors with a scattering surface.
	We use microfacets, which are small, tilted mirrors superimposed on a smooth surface.
	We form a simple model of surface roughness and light scattering based on the orientations of the microfacets.
	Using a least-squares solver to compute the smooth reflector as a starting point, we can subsequently alter the surface using an optimization procedure to account for the scattering.
	After optimization, the resulting reflector surface produces the desired scattered light distribution.
	We verify the resulting reflector using raytracing.
\end{changemargin}
\hrule

\clearpage

\pagestyle{main}
\pagenumbering{arabic}
\setcounter{page}{1}
\restoregeometry

\section{Introduction}\label{sec:intro}
Since a few years, light-emitting diodes (LEDs) have been the dominant mode of artificial lighting worldwide \cite{statista_LED_penetration_rate}.
While LEDs come with several benefits, such as their unrivaled energy efficiency and long lifespan, they are not without downsides.
The one we wish to focus on in this manuscript is the undesirable sharp glare that can occur due to the point-like nature of the LED chip.
Many contemporary solutions to this problem rely on incorporating scattering elements into the design of the luminaire \cite[Sec.~1.8.4]{koshelIlluminationEngineeringDesign2013}.
While this is an effective solution, it requires significant experience from the optical designer \cite[Sec.~1.9]{koshelIlluminationEngineeringDesign2013}.
This is primarily due to the lack of inverse methods, meaning the process relies on an iterative forward process requiring a rich understanding of the effects each alteration will cause that can only come from experience.

This work aims to improve the situation by a proposed algorithm to directly compute the shape of a freeform reflector that solves the so-called \textit{inverse problem of reflector design} with a scattering surface.
Before tackling the scattering version of this problem, let us introduce the relatively well-understood specular one.
Loosely speaking, the specular inverse problem of reflector design seeks the answer to the question ``if this is the incoming light, and this is the desired outgoing light, what should be the shape of the reflector surface?''
In three dimensions, the answer involves solving a so-called Monge-Amp{\`e}re equation, which is a fully nonlinear partial differential equation \cite{prinsInverseMethodsIllumination2014,yadavMongeAmpereProblemsNonquadratic2018,romijnGeneratedJacobianEquations2021}.

Since no actual surface is perfectly smooth, surface light scattering will occur as the light from the source interacts with the reflector.
The loose definition of the inverse problem of reflector design with a scattering surface can thus be ``if this is the incoming light, this is the desired outgoing light, \textit{and} these are the scattering properties of the surface, what should be the shape of the reflector surface?''
In a previous publication, we explored the possibility of transforming the inverse scattering problem to an ``equivalent'' specular one \cite{kronbergThreedimensionalFreeformReflector2023}.
We showed that this can be done successfully under some assumptions, such as an isotropic surface, meaning the scattering properties do not change with position.

There is a significant lack of literature discussing the unification of freeform optics and surface scattering.
The best example we have been able to find is Lin \textit{et al.} \cite{linNovelOpticalLens2015}, who is working with a combination of a freeform scattering and a spherical specular surface to design a lens.
Their approach is based on Bézier curves, where the initial shape was iteratively modified to account for the differences between the prescribed target distribution and the resulting raytraced distribution.

In this manuscript, we instead seek to model the surface roughness using microfacets and compute the shape of the reflector by minimization.
This has several advantages compared to our previous approach, including the possibility of extending the model to non-isotropic surfaces.
We note that microfacets have a rich history in computer-generated imagery (CGI) dating back to the early 80s \cite{cookReflectanceModelComputer1982, pharrPhysicallyBasedRendering2017}.

This manuscript is organized as follows.
First, the proposed algorithm is summarized in Section \ref{sec:algorithm}.
Next, the scattering model based on microfacets is derived in Section \ref{sec:model}.
The minimization procedure is discussed in Section \ref{sec:model:derivation:freeform:minimization}, and the custom raytracer implementing the microfacets used to verify our model is introduced in Section \ref{sec:verification}.
We then show a numerical example to verify our proposed algorithm in Section \ref{sec:examples}.
Finally, Section \ref{sec:conclusions} concludes the manuscript with a summary and discussion of what we covered and some of our future ambitions.

\subsection{Proposed Algorithm}\label{sec:algorithm}
This section summarizes the proposed design algorithm in words.
Given a source distribution, a probability density function for the microfacet orientations, and a desired far-field scattered target distribution:
\begin{enumerate}
	\item Compute the specular reflector shape that transforms the source distribution to the target intensity.
	\item Alter the surface's normals using a minimization procedure to account for the scattering effects.
	\item Compute the final surface from the minimized normals.
\end{enumerate}
The first and last steps can be assumed to be solved problems.
We will show that readily available software packages can realize the second step.
A commercial routine from MathWorks was used in our case, but similar open-source alternatives exist.

Following these steps directly yields a reflector that transforms the source distribution into the desired target distribution when taking scattering into account.
The remainder of this manuscript will detail these steps and showcase how to use the proposed algorithm in practice.

\clearpage
\section{Scattering Model}\label{sec:model}
In this section, we will introduce and derive the microfacet scattering model and apply it in the context of freeform reflector design.

\subsection{Key Assumptions}\label{sec:model:assumptions}
Let us summarize the most important assumptions we shall make throughout our derivation of the microfacet scattering model.
The most fundamental assumption is that light travels in straight lines called \textit{rays} in a medium with a constant refractive index, i.e., we are in the realm of \textit{geometrical optics}.

While not strictly an assumption, we shall utilize \textit{scaled units} throughout this manuscript.
This choice is a natural consequence of the equations being scale-invariant since they can be utilized to design massive reflectors for telescopes on the order of meters down to tiny ones on the order of millimeters.
In illumination applications, the reflectors are typically in the order of centimeters.
In practice, this means we scale the system using the size of our source.
We shall always assume that the source consists of a collimated beam.
Note that this means that it has zero {\'e}tendue.

Another critical assumption is that we can model light scattering within the geometrical optics approximation.
This is not obvious since light surface scattering is inherently a wave phenomenon arising due to phase differences induced by the varying optical pathlengths experienced by light interacting with a microscopically rough surface \cite{harveyLightScatteringCharacteristicsOptical1977}.
Thus, we make the following assumption: \textit{Surface light scattering can be modeled using geometrical optics by considering incoming and outgoing rays, where the outgoing rays follow a probability density related to the surface's bidirectional reflectance distribution function (BRDF).}

Finally, we shall assume that there are no losses in our systems, and we shall restrict our attention to so-called \textit{far-field} problems where the target distribution is an intensity defined in terms of angles.

\subsection{Geometry}\label{sec:model:geometry}
\begin{figure}[hbt!]
	\centering
	\includegraphics{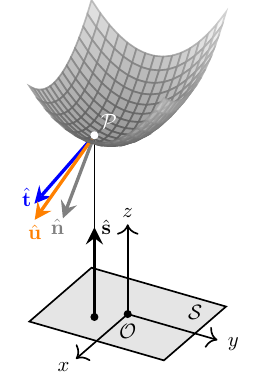}
	\caption{Summary of the geometry used in this manuscript. Note that $\us$, $\un$, and $\ut$ are coplanar in the PoI while $\uu$ generally falls outside of the PoI.}
	\label{fig:general_geometry}
\end{figure}

Before discussing the scattering model in detail, let us consider the general problem geometry (refer to Fig.~\ref{fig:general_geometry}).
Suppose we have a Cartesian $xyz$-coordinate system in $\bbR^3$, with a \textit{parallel source }in the $xy$-plane, i.e., at $z = 0$, represented by the rectangular source domain $\mcS := [a_1,a_2] \times [b_1,b_2] \subset \bbR^2$.
For simplicity, let us align the parallel source rays with the $z$-axis so that if $\us$ is the direction of a source ray, $\us \equiv \ue_z := (0,0,1)^\intercal$.
We shall denote unit vectors with a hat ($\,\hat{\ }\,$) throughout this manuscript.
The reflector is assumed to be above the source domain, and its height is given by $z = u(x,y)$ for all $(x,y) \in \mcS$, where $u(x,y) > 0$ is the (at least) twice-differentiable height function of the reflector.

Let us follow a single source ray, leaving the source at some point in $\mcS$.
It eventually strikes the reflector at some point $\mcP$, where the unit normal of the reflector is given by $\un$.
By convention, we shall orient the normal towards the light source, i.e., $\inner{\us,\un} < 0$.
The \textit{vectorial law of reflection} (LoR) yields the specular direction, denoted $\ut$ herein,
\begin{equation}\label{eq:LoR}
	\ut = \us - 2 \inner{\us,\un} \un.
\end{equation}
Note that $\us$, $\un$, and $\ut$ are coplanar, spanning the so-called \textit{plane of incidence} (PoI).

Suppose instead that the incident light experiences off-specular scattering at $\mcP$.
The outgoing light would then leave the surface in a direction given by the vector $\uu$.
Generally, $\uu$ does not lie in the PoI, and naturally, $\uu \neq \ut$.

\subsection{Model Derivation}\label{sec:model:derivation}
To develop the scattering model, let us start with Nicodemus's work from 1965 \cite{nicodemusDirectionalReflectanceEmissivity1965}.
Specifically, they relate the incident \textit{radiance} $\tLi$ [$\text{W} \cdot \text{m}^{-2} \cdot \text{sr}^{-1}$] to the radiance leaving the surface, $\tLo$.
The relation involves the \textit{bidirectional reflectance distribution function} (BRDF), $\tB$, and holds pointwise on any surface \cite[cf.~Eq.~(7)]{nicodemusDirectionalReflectanceEmissivity1965}:
\begin{equation}\label{eq:Lo_def}
	\tLo(\uu) = \int_{S^2} \tB(\us,\uu) \tLi(\us) \inner{-\us,\un} \, \dd\rmS(\us),
\end{equation}
where $\dd\rmS(\us)$ is a surface area element on the unit sphere depending on the parametrization of $\us$.
The tildes indicate that the radiances and the BRDF take unit (directional) vectors as their arguments.

To use this equation in freeform reflector design, we shall first develop a suitable model for the BRDF of our rough reflector surface.
Any physical BRDF must fulfill the following three conditions \cite{duvenhageNumericalVerificationBidirectional2013}:
\begin{itemize}
	\item Nonnegativity: $\forall \us,\uu \in S^2 : \tB(\us,\uu) \geq 0$.
	\item Helmholtz reciprocity: $\tB(\us,\uu) = \tB(-\uu,-\us)$.
	\item Energy conservation: \begin{equation}\label{eq:BRDF_energyConservation}
		\forall \us, \un \in S^2 : \int_{S^2} \tB(\us,\uu) \inner{\uu,\un} \, \dd\rmS(\uu) = 1.
	\end{equation}
\end{itemize}

\subsubsection{Specular Reflection as a BRDF}\label{sec:model:derivation:specular_BRDF}
As a concrete example of a physical BRDF, let us consider specular reflection.
This section will use the general notation of $\us$ for the incoming direction and $\uu$ as the outgoing direction, even when the latter coincides with the specular one we previously denoted $\ut$.
Recall that specular reflection means that the cosines of the incoming and outgoing angles with respect to the unit normal remain the same.
This is captured in the vectorial law of reflection, Eq.~\eqref{eq:LoR}, which we can invert to get $\us = \uu - 2 \inner{\uu,\un} \un$ since $\inner{-\us, \un} = \inner{\uu, \un}$.
Before defining the specular BRDF, let us introduce an appropriate Dirac delta function for unit vectors on $S^2$.
\begin{definition}\label{def:DiracDelta}
	Let $\uv \in S^2$ be an arbitrary unit vector.
	Then, the Dirac delta function, $\tdelta$, is defined such that
	\begin{equation}\label{eq:DiractDelta_def}
		\int_{S^2} \tdelta(\uv) \, \dd\rmS(\uv) = 1.
	\end{equation}
	Fix $\uv_0 \in S^2$, and let $f$ be an arbitrary function.
	Then, the $\mathrm{sifting\ property}$ of the Dirac delta function reads
	\begin{equation}\label{eq:DiractDelta_sifting}
		\int_{S^2} \tdelta(\uv - \uv_0) f(\uv) \, \dd\rmS(\uv) = f(\uv_0).
	\end{equation}
\end{definition}

\begin{definition}
	Let $\us \in S^2$ be the unit vector indicating the direction of the incoming light, and let $\uu \in S^2$ be the direction of the outgoing light.
	Then, the bidirectional BRDF of specular reflection is given by
	\begin{equation}\label{eq:specularBRDF}
		\tB(\us,\uu) = \frac{\tdelta(\us - \uu + 2\inner{\uu,\un}\un)}{\inner{\uu,\un}},
	\end{equation}
	for a fixed unit normal $\un$.
\end{definition}

\noindent This BRDF fulfills:
\begin{itemize}
	\item Nonnegativity by the properties of the Dirac delta function and the inner product.
	\item Helmholtz reciprocity:\begin{equation}
			\tB(-\uu,-\us) = \frac{\tdelta(-\uu + \us - 2\inner{\us,\un}\un)}{\inner{-\us,\un}} = \tB(\us,\uu),
		\end{equation}
		where the last relation follows directly from $\inner{-\us, \un} = \inner{\uu,\un}$.
	\item Energy conservation: Fix $\us, \un \in S^2$. Then, Eq.~\eqref{eq:BRDF_energyConservation} becomes \begin{equation}\label{eq:BRDF_energyConservation2}
		1 = \int_{S^2} \tB(\us,\uu) \inner{\uu,\un} \, \dd\rmS(\uu) = \int_{S^2} \tdelta(\us - \uu + 2\inner{\uu,\un}\un) \, \dd\rmS(\uu).
	\end{equation}
	Let us now change the integration domain from one parametrized using $\uu$ to one parameterized using $\us$.
	This yields a Jacobian $\abs{J(\uu,\us)}$.
	To evaluate it, let $(\vartheta,\varphi)$ parametrize $\us$, where $\vartheta \in [0,\pi]$ and $\varphi \in [0,2\pi)$ are the polar and azimuth angles in spherical coordinates.
	Using the law of reflection, Eq.~\eqref{eq:LoR}, we can get $\uu(\vartheta,\varphi)$ for fixed $\un$.
	The Jacobian can then be evaluated from the ratio of the area elements $\dd\rmS(\uu)$ and $\dd\rmS(\us)$, i.e.,
	\begin{equation}
		\abs{J(\uu,\us)} = \norm{\pdv{\uu}{\vartheta} \cross \pdv{\uu}{\varphi}}\ \Bigg/\ \norm{\pdv{\us}{\vartheta} \cross \pdv{\us}{\varphi}} = 1.
	\end{equation}
	Finally, for fixed $\un$, let $\us_0(\uu) := \uu - 2\inner{\uu,\un} \un$.
	Then, Eq.~\eqref{eq:BRDF_energyConservation2} can be written as
	\begin{equation}
		1 = \int_{S^2} \tdelta(\us -\us_0) \, \dd\rmS(\us),
	\end{equation}
	which holds by the properties of the Dirac delta function.
\end{itemize}

Inserting the specular BRDF from Eq.~\eqref{eq:specularBRDF} into the relation of Nicodemus, Eq.~\eqref{eq:Lo_def}, yields
\begin{equation}\label{eq:Lo_2}
	\begin{split}
		\tLo(\uu) \inner{\uu,\un} &= \int_{S^2} \tdelta(\us - \uu + 2 \inner{\uu,\un}\un) \tLi(\us) \inner{-\us,\un} \, \dd\rmS(\us)\\
		&= \tLi(\uu - 2\inner{\uu,\un} \un) \inner{-\uu + 2\inner{\uu,\un} \un, \un},
	\end{split}
\end{equation}
where the last relation follows from the law of reflection, Eq.~\eqref{eq:LoR}, and the sifting property of the Dirac delta function, Eq.~\eqref{eq:DiractDelta_sifting}.
By evaluating the inner product and using the law of reflection, we get $\tLo(\uu) = \tLi(\us)$, which is consistent with the fundamental principle of \textit{conservation of radiance} \cite{nicodemusRadiance1963,bunchOpticalSystemsDesign2021}.

\subsubsection{Microfacets}\label{sec:model:derivation:microfacets}
\begin{figure}[hbt!]
	\centering
	\includegraphics[width=0.25\linewidth]{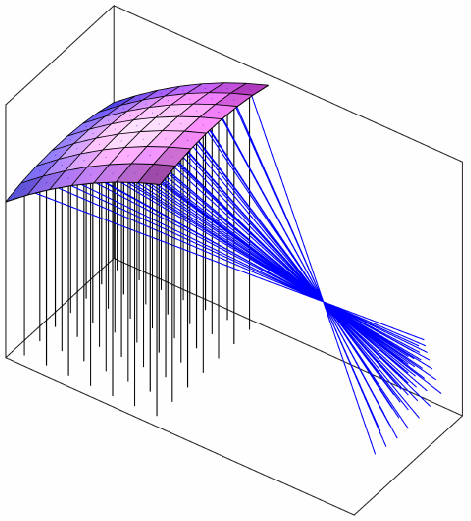}\hspace{7pt}
	\includegraphics[width=0.25\linewidth]{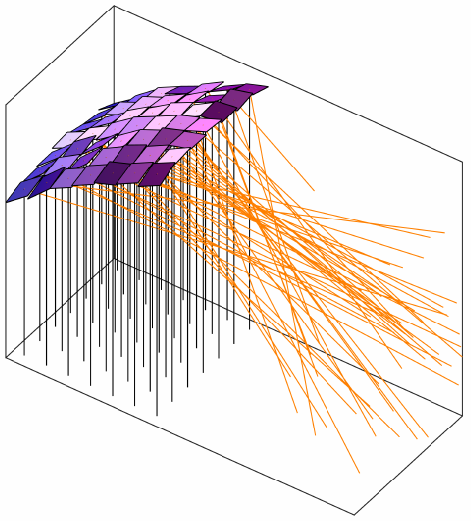}
	\caption{Specular reflection from a smooth surface (left); scattering from a rough surface with microfacets (right).}
	\label{fig:microfacetSchematic}
\end{figure}

We shall use so-called \textit{microfacets} to model surface roughness.
These are small tilted mirrors superimposed on a smooth surface --- see Fig.~\ref{fig:microfacetSchematic} for a schematic representation.
A realistic surface roughness model can be constructed by varying the size and orientations of the microfacets.
This approach has seen extensive use in computer-generated imagery (CGI).
We are particularly interested in the model developed by Torrance and Sparrow in 1967 \cite{torranceTheoryOffSpecularReflection1967} and later applied to CGI by Cook and Torrance in 1982 \cite{cookReflectanceModelComputer1982}.

We have opted to develop a simple model inspired by the one introduced by Torrance and Sparrow.
While this means that our model is less realistic, it lets us focus on our primary goal of unifying surface scattering and inverse reflector design by starting with as few complexities as possible.
We note that the simplicity of our surface roughness model means that it is not straightforward to directly relate it to, e.g., the root mean square (RMS) roughness of an actual surface.

Furthermore, the microfacet distribution function plays a somewhat different role than in CGI since we consider the way in which the microfacet normals are distributed for each ray-reflector intersection point.
Thus, we are at the limit where the microfacets become infinitesimal.
While this is difficult to motivate physically, it makes the model considerably simpler since we can safely omit shadowing and masking effects between neighboring microfacets.
Additionally, this is most similar to the assumptions we made in the optimal transport approach we have previously developed to unify surface scattering and inverse design in two and three dimensions \cite{kronbergModellingSurfaceLight2023,kronbergThreedimensionalFreeformReflector2023}.
Finally, this is analogous to the approach we used for microfacets in two dimensions \cite{kronbergTwodimensionalFreeformReflector2023}.
Extending the microfacets BRDF approach by considering different BRDFs seems like a promising way to improve the physical relevance of this initial model.

\begin{figure}[hbt!]
	\centering
	\includegraphics{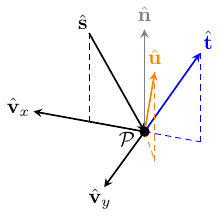}
	\caption{Local geometry of scattering in three dimensions; the PoI contains $\us$, $\un$, $\ut$, and $\uv_x$; $\uv_y$ is perpendicular to the PoI.}
	\label{fig:local_geometry}
\end{figure}

To describe the microfacets, consider zooming in sufficiently close to observe a locally flat area of a freeform reflector --- the squares in Fig.~\ref{fig:microfacetSchematic}.
Recall that $\un$ represents the unit normal of the reflector at some point $\mcP$, and let $\unm$ denote the unit normal of the microfacet at $\mcP$.
Suppose $\unm$ is picked in a cone coaxial with $\un$.
To describe this mathematically, we must first set up the appropriate local geometry close to $\mcP$.
Specifically, we will construct a local Cartesian coordinate system where $\un$ constitutes the local $z$-axis.
The local $x$- and $y$-axes will be represented by two unit vectors, $\uv_x$ and $\uv_y$, where we choose $\uv_x$ in the PoI --- see Fig.~\ref{fig:local_geometry}.

Let $\kappa_1 \in [0,\pi]$ and $\kappa_2 \in [0, 2 \pi)$ be spherical polar and azimuth angles parametrizing $\un$.
Clearly, the local $y$-axis should be perpendicular to the PoI, i.e., to $\us \equiv \ue_z$ and $\un$:
\begin{equation}
	\uv_y := \frac{\ue_z \cross \un(\kappa_1,\kappa_2)}{\norm{\ue_z \cross \un(\kappa_1,\kappa_2)}} =
	\begin{pmatrix}
		-\sin(\kappa_2)\\
		\cos(\kappa_2)\\
		0
	\end{pmatrix}.
\end{equation}
The local $x$-axis should be perpendicular to both $\un$ and $\uv_y$, i.e., in the plane of incidence, e.g.,
\begin{equation}
	\uv_x := \uv_y \cross \un(\kappa_1,\kappa_2) =
	\begin{pmatrix}
		\cos(\kappa_1) \cos(\kappa_2)\\
		\cos(\kappa_1) \sin(\kappa_2)\\
		-\sin(\kappa_1)
	\end{pmatrix}.
\end{equation}
Note that the local $\uv_x$, $\uv_y$, $\un$ coordinate system is right-handed.

\paragraph{Rodrogues' Rotation Matrix}
Let us now define the appropriate rotation matrices to construct $\unm$.
Let $\bfI$ be the $3 \times 3$ identity matrix, and suppose we want to define a rotation matrix around an axis given by the arbitrary unit vector $\uk := (k_1, k_2, k_3)^\intercal \in S^2$.
Rodrigues' rotation matrix $\bfR_{\uk}(\theta)$ that defines a right-handed rotation of angle $\theta$ around $\uk$ is then given by \cite[Eq.~(15)]{daiEulerRodriguesFormula2015}
\begin{subequations}\label{eq:Rodrigues_def}
	\begin{align}
		\bfR_{\uk}(\theta) &:= \bfI + \sin(\theta) \bfK + \big(1 - \cos(\theta)\big) \bfK^2,\\[5pt]
		\bfK &:= \begin{pmatrix}
			0 & -k_3 & k_2\\
			k_3 & 0 & -k_1\\
			-k_2 & k_1 & 0
		\end{pmatrix},
	\end{align}
\end{subequations}
where $\bfK$ is called the \textit{cross-product matrix} since $\bfK \bfa = \uk \cross \bfa$ for an arbitrary vector $\bfa \in \bbR^3$.

\begin{figure}[hbt!]
	\centering
	\includegraphics{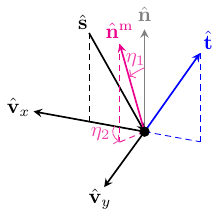}
	\caption{Construction of the mirofacet normal.}
	\label{fig:microfacetNormal}
\end{figure}

We are now ready to construct $\unm$.
Let $\eta_1$ and $\eta_2$ be the polar and azimuth angles of $\unm$ in the local coordinate system --- see Fig.~\ref{fig:microfacetNormal}.
Then,
\begin{equation}\label{eq:unm_def}
	\unm := \bfR_{\un}(\eta_2) \bfR_{\uv_y}(\eta_1) \un,
\end{equation}
where $\bfR_{\uv_y}$ and $\bfR_{\un}$ are Rodrigues rotation matrices defined in accordance with Eq.~\eqref{eq:Rodrigues_def}.
 Note that, by construction,
\begin{subequations}\label{eq:eta1_eta2}
 	\begin{align}
 		\cos(\eta_1) &= \inner{\unm,\un},\label{eq:eta_1}\\
 		\tan(\eta_2) &= \frac{\inner{\unm,\uv_y}}{\inner{\unm,\uv_x}}.
	\end{align}
\end{subequations}
 
\paragraph{Distributed Orientations}
To use microfacets to describe scattering from a surface, rather than fixing $\eta_1$ and $\eta_2$, we shall consider that they are sampled from some distribution related to the roughness properties of the surface.
That is, we do not know \textit{a priori} which microfacet orientation we strike, but instead, we wish to describe the probability density function (PDF) $p(\eta_1,\eta_2)$ with support $\supp(p) = [0,\pi] \times [0, 2\pi)$, i.e.,
\begin{equation}
	\int_{0}^{2\pi} \int_{0}^{\pi} p(\eta_1,\eta_2) \sin(\eta_1) \, \dd\eta_1 \, \dd\eta_2 = 1.
\end{equation}
Using the relations in Eq.~\eqref{eq:eta1_eta2}, we can also write, for all $\un \in S^2$,
\begin{equation}
	\int_{S^2} p\left(\arccos(\inner{\unm,\un}), \arctan(\inner{\unm,\uv_x},\inner{\unm,\uv_y})\right) \dd\rmS(\unm) = 1,
\end{equation}
where $\arctan(x,y)$ is the arc tangent of $y/x$, taking into account the quadrant of the point $(x,y)$, and $\dd\rmS(\unm)$ represents an area element on the unit sphere generated by $\unm$.
While the support of $p$ can formally extend to $[0,\pi] \times [0, 2\pi)$, we will typically consider situations where the significant contributions in the polar direction are concentrated in a narrow region, peaking at $\eta_1 = 0$.
This shape is consistent with measured BRDFs from machined surfaces \cite[Ch.~4]{stoverOpticalScatteringMeasurement2012}.

To get the outgoing radiance, we now multiply the right-hand side of Eq.~\eqref{eq:Lo_2} by $p$ and integrate over all valid microfacet orientations, leading to the definition below.
\begin{definition}
	Let $\eta_1 \in [0,\pi]$ and $\eta_2 \in [0,2\pi)$ be the polar and azimuth angles of the microfacet in the local coordinate system, sampled from the probability density function $p(\eta_1, \eta_2)$.
	Let $\tLo$ be the outgoing (i.e., scattered) radiance in direction $\uu$ due to an incoming radiance $\tLi$ in direction $\us$.
	Then:
	\begin{equation}\label{eq:Lo_general}
		\begin{split}
			\tLo(\uu) \inner{\uu,\un} = \int_{S^2} &p\!\left(\arccos(\inner{\unm,\un}), \arctan(\inner{\unm,\uv_x},\inner{\unm,\uv_y})\right)\\
			& \tLi(\uu - 2\inner{\uu,\unm} \unm) \inner{-\uu + 2\inner{\uu,\unm} \unm, \un} \, \dd\rmS(\unm).
		\end{split}
	\end{equation}
\end{definition}

\paragraph{Rotationally Symmetric Scattering}
Under the assumption of rotationally symmetric scattering, i.e., when $p(\eta_1,\eta_2) = p(\eta_1)$ for all $\eta_2 \in [0,2\pi)$, the normalization of $p$ becomes
\begin{equation}\label{eq:energyConservation_p_symmetric}
	2\pi \int_{0}^{\pi} p(\eta_1) \, \sin(\eta_1) \, \dd\eta_1 = 1,
\end{equation}
and Eq.~\eqref{eq:Lo_general} reads
\begin{equation}\label{eq:Lo_rotSym}
	\begin{split}
		\tLo(\uu) \inner{\uu,\un} = \int_{S^2} &p\big(\!\arccos(\unm,\un)\big) \tLi\big(\uu - 2\inner{\uu,\unm}\unm\big) \\
		&\big\langle\! -\uu + 2\inner{\uu,\unm}\unm,\un \big\rangle \, \dd\rmS(\unm),
	\end{split}
\end{equation}
where $\arccos(\unm,\un)$ is shorthand for $\arccos(\inner{\unm,\un})$.

\begin{figure}[hbt!]
	\centering
	\includegraphics{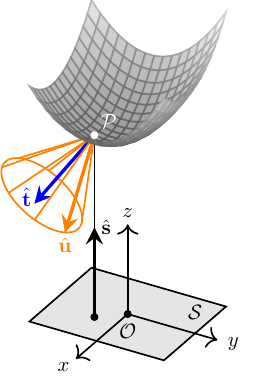}
	\caption{Rotationally symmetric scattering implies that $\uu$ is equally likely to be anywhere on a cone coaxial with $\ut$ for fixed $\eta_1$.}
	\label{fig:uu_sym}
\end{figure}

Note that $\eta_2$ being uniform on $[0,2\pi)$ means that $\unm$ is equally likely to be anywhere on a cone coaxial with $\un$ for fixed $\eta_1$.
This translates to $\uu$ being equally likely to be anywhere on a cone coaxial with $\ut$ for fixed $\eta_1$ via the LoR, Eq.~\eqref{eq:LoR}, applied to the microfacet --- see Fig.~\ref{fig:uu_sym}.
We shall focus on the rotationally symmetric case henceforth for simplicity.

\subsubsection{Freeform Reflector Design}\label{sec:model:derivation:freeform}
Note that Eq.~\eqref{eq:Lo_rotSym} holds pointwise on any surface.
Let $\bfr \in \mcR \subset \bbR^3$ be a point on the surface $\mcR$ representing a freeform reflector.
Then, for a surface that may be anisotropic, in the sense that $p$ changes with position, Eq.~\eqref{eq:Lo_rotSym} becomes,
\begin{equation}
	\begin{split}
		\tLo(\bfr,\uu) \inner{\uu,\un} = \int_{S^2} &p\big(\bfr,\arccos(\unm,\un)\big) \tLi\big(\bfr,\uu - 2\inner{\uu,\unm}\unm\big)  \\
		&\big\langle\! -\uu + 2\inner{\uu,\unm}\unm,\un \big\rangle \, \dd\rmS(\unm).
	\end{split}
\end{equation}
In the far field, we observe all the light leaving the reflector, so we must integrate over $\mcR$, keeping in mind that $\un$ changes with position, i.e.,
\begin{equation}\label{eq:hGamma1}
	\begin{split}
		&\int_{\mcR} \tLo(\bfr,\uu) \inner{\uu,\un(\bfr)} \, \dd \rmS(\bfr) =\\
		& \int_{\mcR}\int_{S^2} p\big(\bfr,\arccos(\unm,\un(\bfr))\big) \tLi\big(\bfr,\uu - 2\inner{\uu,\unm}\unm\big) \\
		&\qquad \quad \ \big\langle\! -\uu + 2\inner{\uu,\unm}\unm,\un(\bfr) \big\rangle \, \dd\rmS(\unm) \, \dd \rmS(\bfr),
	\end{split}
\end{equation}
where $\dd \rmS(\bfr)$ is an area element on the reflector surface.
Notice that the left-hand side only depends on $\uu$.
Let us take this as the definition of $\tilde{h}(\uu)$, which represents the \textit{scattered intensity} [$\textrm{W} \cdot \textrm{sr}^{-1}$] in direction $\uu$.

To proceed, note that $\unm$ is the normalized bisector between $-\us$ and $\uu$ due to the LoR, Eq.~\eqref{eq:LoR}, applied to the microfacet.
That is,
\begin{equation}\label{eq:microfacetBisector}
	\unm = \frac{\uu - \us}{\norm{\uu - \us}}.
\end{equation}
This allows us to write Eq.~\eqref{eq:hGamma1} in terms of $\us$, $\uu$ and $\un$ using $\us = \uu - 2\inner{\uu,\unm}\unm$, i.e.,
\begin{equation}\label{eq:hGamma_Li}
	\begin{split}
		\tilde{h}(\uu) = \int_{\mcR}\int_{S^2} &p\left(\bfr,\arccos(\frac{\uu - \us}{\norm{\uu - \us}},\un(\bfr))\right) \tLi\big(\bfr,\us\big) \big\langle\! -\us,\un(\bfr)\big\rangle \\
		& \abs{J(\unm,\us)} \, \dd\rmS(\us) \, \dd \rmS(\bfr),
	\end{split}
\end{equation}
where $\abs{J(\unm,\us)}$ is the Jacobian associated with the change of integration from $\unm$ to $\us$.

Next, let us utilize the fact that we are restricting our attention to parallel sources to simplify the expression for $\tilde{h}$.
Specifically, the radiance of a parallel source with \textit{emittance} $f(\bfx) > 0$ [$\textrm{W} \cdot \textrm{m}^{-2}$] at $\bfx \in \mcS$ is given by
\begin{equation}
	\tLi\big(\bfr,\us\big) := f(\bfx(\bfr)) \, \tdelta(\us - \ue_z).
\end{equation}

\paragraph{Isotropic surfaces}
We shall restrict our attention to isotropic surfaces henceforth, meaning the explicit dependence on $\bfr$ in $p$ can be dropped since the scattering function is the same everywhere.
Then, Eq.~\eqref{eq:hGamma_Li} becomes
\begin{equation}
	\begin{split}
		\tilde{h}(\uu) &= \int_{\mcR}\int_{S^2} p\left(\!\arccos(\frac{\uu - \us}{\norm{\uu - \us}},\un(\bfr))\right) f(\bfx(\bfr)) \tdelta(\us - \ue_z) \\
			&\qquad \qquad \big\langle\! -\us,\un(\bfr)\big\rangle \, \abs{J(\unm,\us)} \, \dd\rmS(\us) \, \dd \rmS(\bfr)\\
		&= \int_{\mcR} p\left(\!\arccos(\frac{\uu - \ue_z}{\norm{\uu - \ue_z}},\un(\bfr))\right) f(\bfx(\bfr)) \big\langle\! -\ue_z,\un(\bfr)\big\rangle \\
		&\qquad \quad \abs{J(\unm,\us)}\eval_{\us = \ue_z} \, \dd \rmS(\bfr)\\
		&= \int_{\mcS} p\left(\!\arccos(\frac{\uu - \ue_z}{\sqrt{2}\sqrt{1 - u_3}},\un(\bfr(\bfx)))\right) f(\bfx) \\
		&\qquad \quad \abs{J(\unm,\us)}\eval_{\us = \ue_z} \, \dd \rmA(\bfx),
	\end{split}
\end{equation}
where we moved to integration over the parallel source domain in the last step, and where $u_3$ is the third component of $\uu$.
Note that
\begin{equation}
	\dd\rmS(\bfr) = \sqrt{1 + \norm{\grad u(\bfx)}^2} \, \dd\rmA(\bfx) = -\frac{1}{n_3} \, \dd\rmA(\bfx),
\end{equation}
where $\dd\rmA(\bfx) = \dd x \, \dd y$ is an area element on the source.

To proceed, we should find an expression for the Jacobian.
First, fix $\uu \in S^2$.
Then, parametrize $\us$ using polar and azimuth angles, $\vartheta$ and $\varphi$.
Since $\unm$ is the normalized bisector between $-\us$ and $\uu$, we can use Eq.~\eqref{eq:microfacetBisector} to get $\unm(\vartheta,\varphi)$.
The Jacobian we are looking for can now be evaluated (with significant effort; we employed Mathematica) using the ratio of the area elements $\dd\rmS(\unm)$ and $\dd\rmS(\us)$, i.e.,
\begin{equation}\label{eq:JacobianFucked_s}
	\abs{J(\unm,\us)} = \norm{\pdv{\unm}{\vartheta} \cross \pdv{\unm}{\varphi}}\ \Bigg/\ \norm{\pdv{\us}{\vartheta} \cross \pdv{\us}{\varphi}} = \frac{1}{2 \sqrt{2} \sqrt{1 - \inner{\us,\uu}}}.
\end{equation}
Hence,
\begin{equation}
	\abs{J(\unm,\us)}\eval_{\us = \ue_z} = \frac{1}{2 \sqrt{2} \sqrt{1 - u_3}}.
\end{equation}

\noindent This leaves us with
\begin{equation}
	\tilde{h}(\uu) = \frac{1}{2 \sqrt{2} \sqrt{1 - u_3}} \int_{\mcS} p\left(\!\arccos(\frac{\uu - \ue_z}{\sqrt{2}\sqrt{1 - u_3}},\un(\bfr(\bfx)))\right) f(\bfx) \, \dd \rmA(\bfx).
\end{equation}
Suppose $\uu$ and $\un$ are parametrized using spherical polar angles $\gamma$, $\kappa_1$ and azimuth angles $\nu$, $\kappa_2$, respectively.
Let $h(\gamma,\nu) := \tilde{h}(\uu(\gamma,\nu))$ so that
\begin{equation}\label{eq:microfacetScatterEq_parallel}
	\begin{split}
		h(\gamma,\nu) = \frac{1}{2 \sqrt{2} \sqrt{1 - \cos(\gamma)}} \int_{b_1}^{b_2} \int_{a_1}^{a_2} &p\left(\eta_1(\gamma,\nu,\kappa_1(x,y),\kappa_2(x,y))\right) \\
		&f(x,y) \, \dd x \, \dd y,
	\end{split}
\end{equation}
where we have slightly abused the notation by reusing $\kappa_1$ and $\kappa_2$ as functions of $\bfx = (x,y)^\intercal$ on the source rather than of $\bfr$ on the reflector.
The polar angle of the microfacet normal in the local coordinate system becomes (recall Eqs.~\eqref{eq:unm_def} and \eqref{eq:eta_1})
\begin{equation}
	\begin{split}
		&\eta_1(\gamma,\nu,\kappa_1,\kappa_2) :=\\
		&\arccos(\frac{(\cos(\gamma) - 1)\cos(\kappa_1) + \sin(\gamma) \sin(\kappa_1) \cos(\nu - \kappa_2)}{2\sin(\gamma/2)}).
	\end{split}
\end{equation}

\subsubsection{Minimization}\label{sec:model:derivation:freeform:minimization}
Our proposed solution algorithm to compute a freeform reflector surface with surface scattering that produces a desired far-field target intensity is based on minimization.
More specifically, we shall optimize the normal $\un$ along the reflector so that the integral in Eq.~\eqref{eq:microfacetScatterEq_parallel} approaches the prescribed scattered target distribution, $h$.
The procedure that finds the optimal solution in a least-squares sense can be summarized as follows:
\begin{enumerate}
	\item Compute the reflector that solves the specular problem of transforming $f$ into $h$.
	\item Compute $\kappa_1(x,y)$ and $\kappa_2(x,y)$, i.e., the polar and azimuth angles of the unit normal $\un$ at each point $(x,y) \in \mcS$. This will be our initial guess $\boldsymbol{\kappa}_0(x,y) := (\kappa_{0,1}(x,y), \kappa_{0,2}(x,y))^\intercal$.
	\item Find \begin{equation} \label{eq:toMinimize} \begin{split}
			\min_{\kappa_1(x,y),\, \kappa_2(x,y)} &\int_{0}^{2\pi} \int_{0}^{\pi} \Bigg( h(\gamma,\nu) - \frac{1}{2 \sqrt{2} \sqrt{1 - \cos(\gamma)}} \\
				&\int_{b_1}^{b_2} \int_{a_1}^{a_2} p\left(\eta_1(\gamma,\nu,\kappa_1(x,y),\kappa_2(x,y))\right) \\
				&\qquad\qquad\quad f(x,y) \, \dd x \, \dd y \Bigg)^2 \sin(\gamma) \, \dd\gamma \, \dd\nu,
			\end{split}
		\end{equation}
		starting with $\kappa_1(x,y) = \kappa_{0,1}(x,y)$ and $\kappa_2(x,y) = \kappa_{0,2}(x,y)$.
	\item Compute the reflector surface from the minimized normals.
\end{enumerate}
The first and last steps can be considered solved problems from our perspective.
Nevertheless, a short summary of the procedures is available below.

\paragraph{Inverse Specular Problem}
Regarding the computation of the reflector surface, we must first address how the initial, specular reflector was computed.
To this end, we used the numerical least-squares Monge-Amp{\`e}re solver first introduced in our group by Prins \cite{prinsInverseMethodsIllumination2014} and later expanded by Yadav \cite{yadavMongeAmpereProblemsNonquadratic2018} and Romijn \cite{romijnGeneratedJacobianEquations2021}.
The details are outside the scope of this work, but a summary is given below.

The problem of computing the reflector reduces to solving a fully nonlinear partial differential equation (PDE).
Let $\hat{h}$ be the stereographic representation of $h(\gamma,\nu)$ such that $\hat{h}(\bfy) = h\big(\gamma(\bfy),\nu(\bfy)\big)$, where $\bfy$ is the 2-tuple stereographic representation of $\uu$ defined according to
\begin{equation}\label{eq:sterNP}
	\bfy(\uu) =
	\begin{pmatrix}
		y_1\\
		y_2
	\end{pmatrix}
	=
	\frac{1}{1-u_3}
	\begin{pmatrix}
		u_1\\
		u_2
	\end{pmatrix}
	=
	\frac{\sin(\gamma)}{1-\cos(\gamma)}
	\begin{pmatrix}
		\cos(\nu)\\
		\sin(\nu)
	\end{pmatrix}.
\end{equation}
By using energy conservation and and $\bfy = \grad u$, we recover the \textit{standard Monge-Amp{\`e}re equation} \cite[Sec.~3.2]{romijnGeneratedJacobianEquations2021}
\begin{equation}\label{eq:MongeAmpere}
	\det(\rmD^2 u(\bfx)) = \frac{1}{4} \big(1 + \norm{\grad u(\bfx)}^2\big)^2 \frac{f(\bfx)}{\hat{h}\big( \grad u(\bfx) \big)},
\end{equation}
where $\rmD^2 u(\bfx)$ denotes the Hessian matrix.
To find the reflector, we must solve the Monge-Amp{\`e}re equation for the specular height function $u$, subject to the so-called \textit{transport boundary condition}, which ensures that all the light from the source reaches the target \cite{riesEdgerayPrincipleNonimaging1994}, \cite[Sec.~6.2.1]{prinsInverseMethodsIllumination2014}.

\paragraph{Reflector from Normals}
For the final step in the procedure, we must compute the reflector surface from its minimized normals.
To find the surface height function, $u$, in a least-squares sense, we can use
\begin{equation}\label{eq:surfaceFunctional}
	I[u] = \frac{1}{2} \int_{\mcS} \norm{\grad u(\bfx) - \bfm(\bfx)}^2 \, \dd\rmA(\bfx),
\end{equation}
where the mapping, $\bfm$, can be expressed via the components of $\un = (n_1,n_2,n_3)^\intercal$ as
\begin{equation}
	\bfm :=
	 -\frac{1}{n_3}
	\begin{pmatrix}
		n_1\\
		n_2
	\end{pmatrix}.
\end{equation}
Note that we get the unit normal from the minimization procedure via $\un = \un(\kappa_1,\kappa_2)$.

To proceed, we must compute the first variation of the functional in Eq.~\eqref{eq:surfaceFunctional} and use Gauss's theorem together with the fundamental lemma of the calculus of variations.
This results in a Neumann problem, which is then solved for $u$.
For more details, including how to make the solution unique, we refer the reader to \cite[Sec.~6.1.4]{romijnGeneratedJacobianEquations2021}.

\paragraph{Numerical Solution Algorithm}
We shall now briefly summarize the numerical version of the proposed procedure:
\begin{enumerate}
	\item Discretize the source and target domains, $\mcS$ and $\mcU$, using $(N_1 + 1) \times (N_2 + 1)$ rectangular grids.
	\item Compute the discretized source and target distributions by evaluating $f$ and $h$ at the $N_1 \times N_2$ center points of the grids. Call them $\mathbf{f}$ and $\mathbf{h}$.
	\item Compute the discretized reflector that solves the specular problem of transforming $\mathbf{f}$ to $\mathbf{h}$.
	\item Compute the unit normals of the reflector and their associated spherical angles. Call the tensor containing the spherical angles at each point $\boldsymbol{\kappa}_0$.
	\item Find the minimum of Eq.~\eqref{eq:toMinimize} using Matlab's \texttt{fminunc} routine and numerically approximate the integrals.
	\item Numerically solve the Neumann problem to get the reflector height after optimizing the normals.
\end{enumerate}

\section{Verification}\label{sec:verification}
To verify the computed reflector shapes, we raytraced the resulting reflectors using a custom raytracer, which implements our microfacets scattering model.

\subsection{Raytracer}
Suppose we have computed the freeform reflector on an $N_1 \times N_2$ rectangular grid in the source domain $\mcS$.
Firstly, using Matlab's \texttt{surfnorm} routine, we can get the surface's $N_1 \times N_2$ normals.
By convention, we shall orient the normals toward the source, i.e., with a negative $z$-component.

Suppose now a source ray travels along $\us \equiv \ue_z$.
To pick the location of the source ray, we sample a point $\bfx \in \mcS$.
Next, we find the normal $\un$ at the point of intersection between the source ray and the reflector surface using Matlab's \texttt{interp2} command.
This routine performs two-dimensional interpolation via table lookup, and we used the \texttt{linear} option.
We thus assume that the unit normal changes linearly in each dimension between its neighboring unit normals.

Once we have determined the unit normal at the intersection point, $\un$, we compute its spherical angles $\kappa_1$ and $\kappa_2$.
We then sample $\eta_1$ from the prescribed microfacet probability density function, $p$ (satisfying Eq.~\eqref{eq:energyConservation_p_symmetric}), and $\eta_2$ uniformly on $[0,2\pi)$.
This allows us to compute $\unm$ using Eq.~\eqref{eq:unm_def}.
The vectorial law of reflection, Eq.~\eqref{eq:LoR}, for the microfacet then gives the scattered direction, i.e., $\uu = \us - 2\inner{\us,\unm}\unm$.
The sampling of $\eta_1$ and $\eta_2$ will be discussed in detail below.

Now that we know the origin of $\us$ and the direction of the scattered ray, $\uu$, we must collect them to compare the result to the prescribed distributions.
Suppose the source and target domains are discretized into $N_{\rmb_1} \times N_{\rmb_2}$ equispaced rectangles, so-called \textit{bins}.
The correct bin for each ray is then identified using Matlab's \texttt{dsearchn} nearest point search function, and the number of rays counted in the returned bin is incremented by unity.
To compare the ray counts to the prescribed distributions, we must convert them to an exitance $f$ (for the parallel source) or an intensity $h$ (for the target).
The relations are as follows, for all $i = 1,2,\dots,N_{\rmb_1}$ and $j = 1,2,\dots,N_{\rmb_2}$:
\begin{subequations}\label{eq:raytracing_intensity}
	\begin{align}
		f_{ij} &= \frac{\mathrm{Pr}(x_{i-1} \leq x < x_i \ \wedge \ y_{j-1} \leq y < y_j)}{\Delta x \, \Delta y} \, \Phi_\rms,\\
		h_{ij} &= \frac{\mathrm{Pr}(\gamma_{i-1} \leq \gamma < \gamma_i \ \wedge \ \nu_{i-1} \leq \nu < \nu_i)}{\sin(\gamma_i) \, \Delta \gamma \, \Delta \nu} \, \Phi_\rms,
	\end{align}
\end{subequations}
where $\Phi_\rms$ is the flux of the source, i.e.,
\begin{equation}
	\Phi_{\rms} := \int_{\mcS} f(x,y) \, \dd x \, \dd y,
\end{equation}
and where $\mathrm{Pr}(x_{i-1} \leq x < x_i \ \wedge \ y_{j-1} \leq y < y_j)$ or $\mathrm{Pr}(\gamma_{i-1} \leq \gamma < \gamma_i \ \wedge \ \nu_{i-1} \leq \nu < \nu_i)$ is the number of rays in the $ij$th bin divided by the total number of rays traced, i.e., the probability falling in the $ij$th bin.
Finally, $\Delta x \, \Delta y$ or $\sin(\gamma_i) \, \Delta \gamma \, \Delta \nu$ is the size of each bin.

\paragraph{Sampling of $\eta_1$ and $\eta_2$.}
We shall now consider the problem of sampling $\eta_1$ and $\eta_2$ in our raytracer.
First, we pick two independent normally distributed variables $q_1 \sim \mcN(0,\sigma)$ and $q_2 \sim \mcN(0,\sigma)$, where $\mcN(\mu,\sigma)$ is the normal distribution with mean $\mu$ and standard deviation $\sigma$.
Consequently, the point $\bfq := (q_1,q_2)^\intercal \in \mathbb{R}^2$ is picked from a rotationally symmetric two-dimensional normal distribution.
Applying inverse stereographic projection towards the south pole from $\bfq$ thus yields a point on the unit sphere.
Note that we chose the south pole since we want the peak of the Gaussian in the stereographic plane to be at $\bfq = \boldsymbol{0}$, representing $\eta_1 = 0$, where the stereographic projection from the north pole is undefined.
Let $\uv$ be the resulting unit vector after applying the inverse stereographic projection.
Then,
\begin{equation}
	\uv = \frac{1}{1+\norm{\bfq}^2}
	\begin{pmatrix}
		2q_1\\
		2q_2\\
		1-\norm{\bfq}^2\\
	\end{pmatrix}.
\end{equation}
This allows us to find closed expressions for $\eta_1$ and $\eta_2$ in terms of $\bfq$:
\begin{equation}\label{eq:sampleRT}
	\begin{split}
		\eta_1(q_1,q_2) &= \arccos\left(\frac{1 - \norm{\bfq}^2}{1+\norm{\bfq}^2} \right),\\
		\eta_2(q_1,q_2) &= \arctan(q_1,q_2).
	\end{split}
\end{equation}
It can be shown that the PDF, $p(\eta_1; \sigma)$, associated with this approach of picking $\eta_1$ is given by \cite[Appendix A]{kronbergThreedimensionalFreeformReflector2023}
\begin{equation}\label{eq:pAlpha}
	p(\eta_1; \sigma) = \frac{1}{8\pi \sigma^2} \, \sec^4\!\bigg(\frac{\eta_1}{2}\bigg) \exp\!\Bigg(-\frac{1}{2\sigma^2} \tan^2\!\bigg(\frac{\eta_1}{2}\bigg)\Bigg).
\end{equation}
Whence, the raytracer picks $\eta_1$ and $\eta_2$ by sampling $q_1 \sim \mcN(0,\sigma)$ and $q_2 \sim \mcN(0,\sigma)$ followed by applying Eq.~\eqref{eq:sampleRT}.
The predicted scattered light distribution, meanwhile, is computed by inserting $p$ from Eq.~\eqref{eq:pAlpha} into Eq.~\eqref{eq:microfacetScatterEq_parallel} with known $f$ and $h$.

\subsection{RMS Error}
In addition to qualitatively comparing the raytraced distributions with the prescribed ones, we shall utilize the root mean square (RMS) error for a quantitative comparison.
The RMS error between the prescribed target distribution $h$ and the raytraced distribution $h^*$ with $N_{\mathrm{b}_1} \times N_{\mathrm{b}_2}$ bins is defined as
\begin{equation}\label{eq:RMS}
	\varepsilon(h,h^*) := \sqrt{ \frac{1}{N_{\rmb_1} N_{\rmb_2}} \sum_{j=1}^{N_{\rmb_2}} \sum_{i=1}^{N_{\rmb_1}} \abs{h_{ij} - h^*_{ij}}^2 }.
\end{equation}
Note that an upper-index asterisk $({}^{*})$ denotes a raytraced distribution in the following section.

\section{Numerical Example}\label{sec:examples}
\begin{figure*}[htb!]
	\centering
	\includegraphics[width=0.33\linewidth]{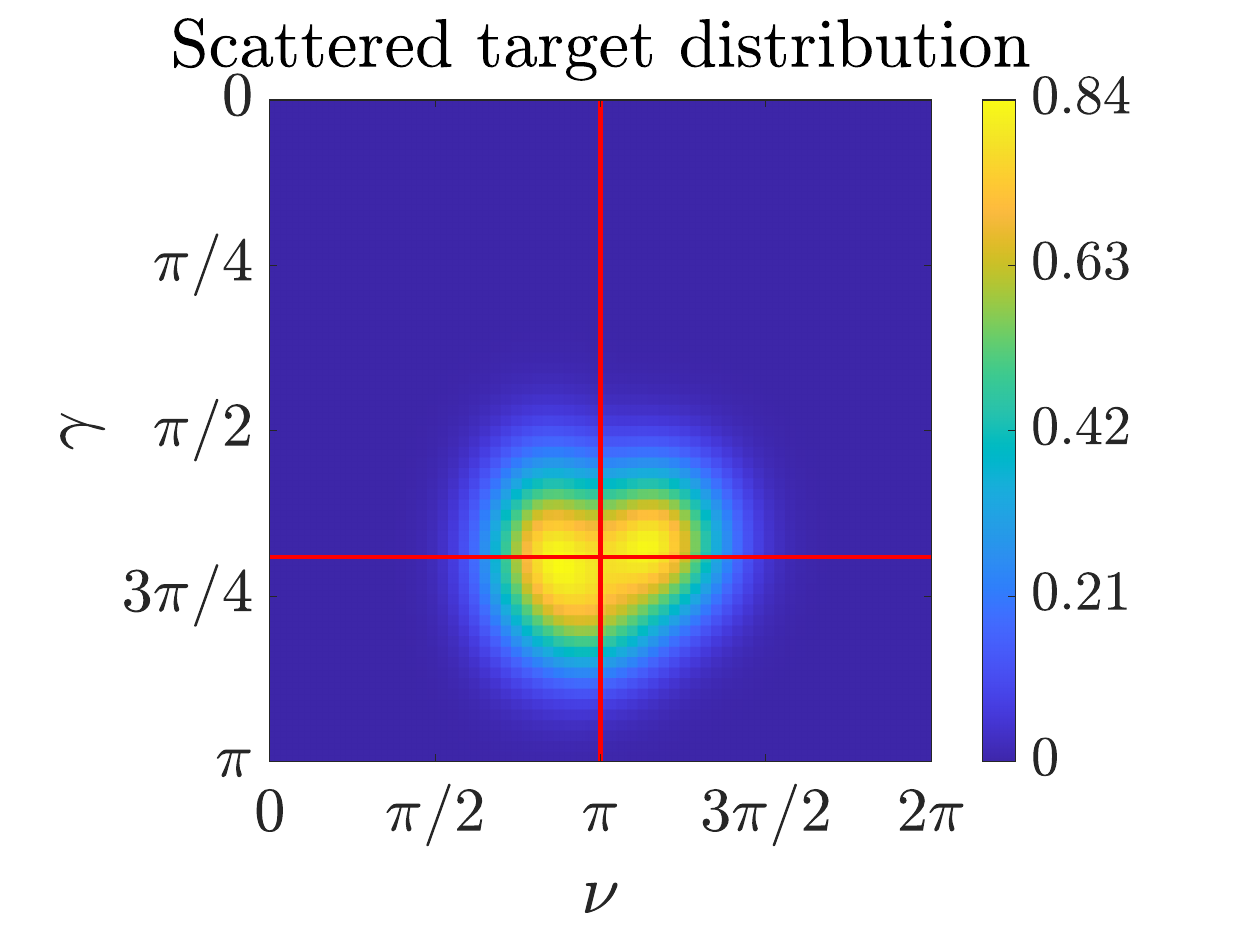}\hfill
	\includegraphics[width=0.33\linewidth]{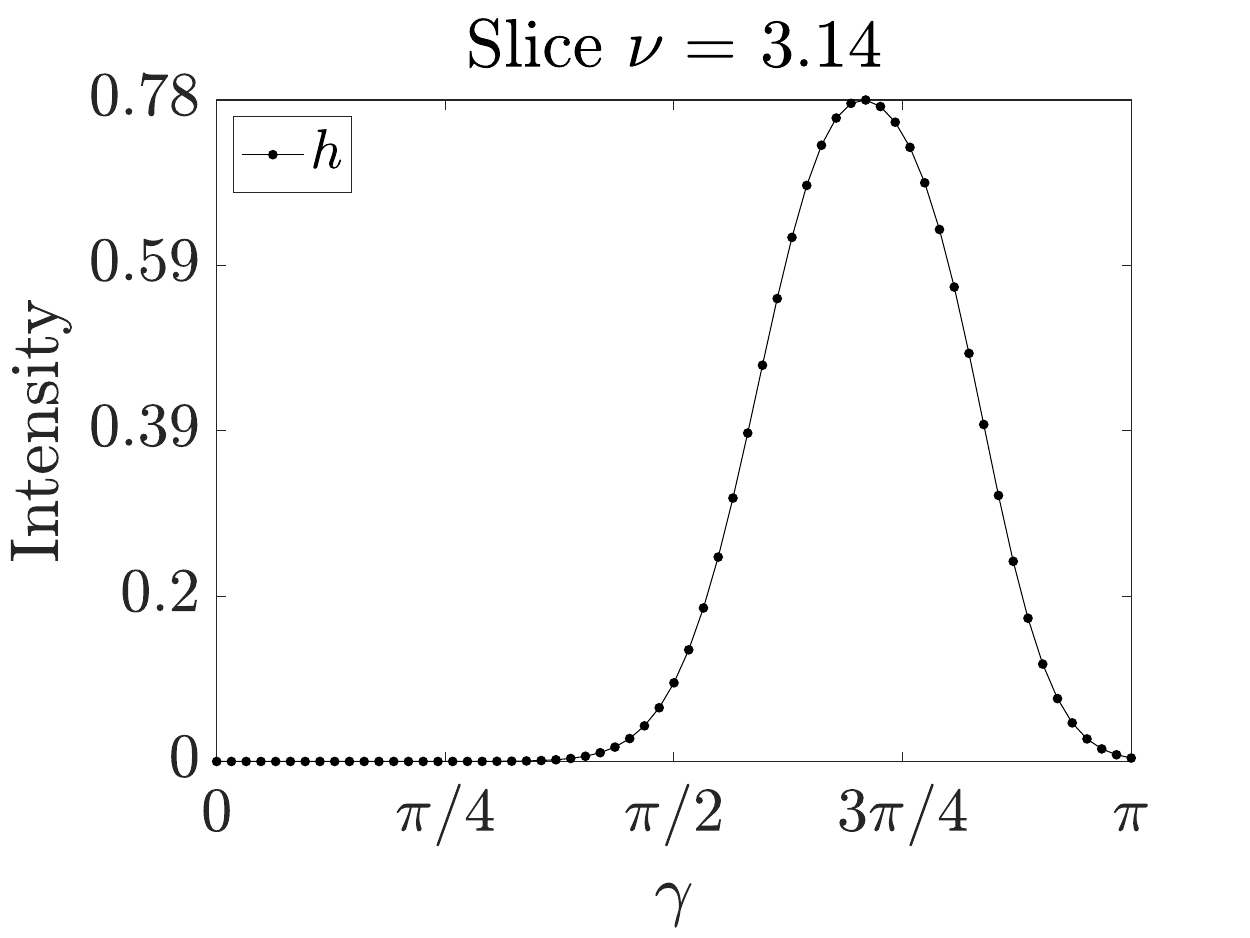}\hfill
	\includegraphics[width=0.33\linewidth]{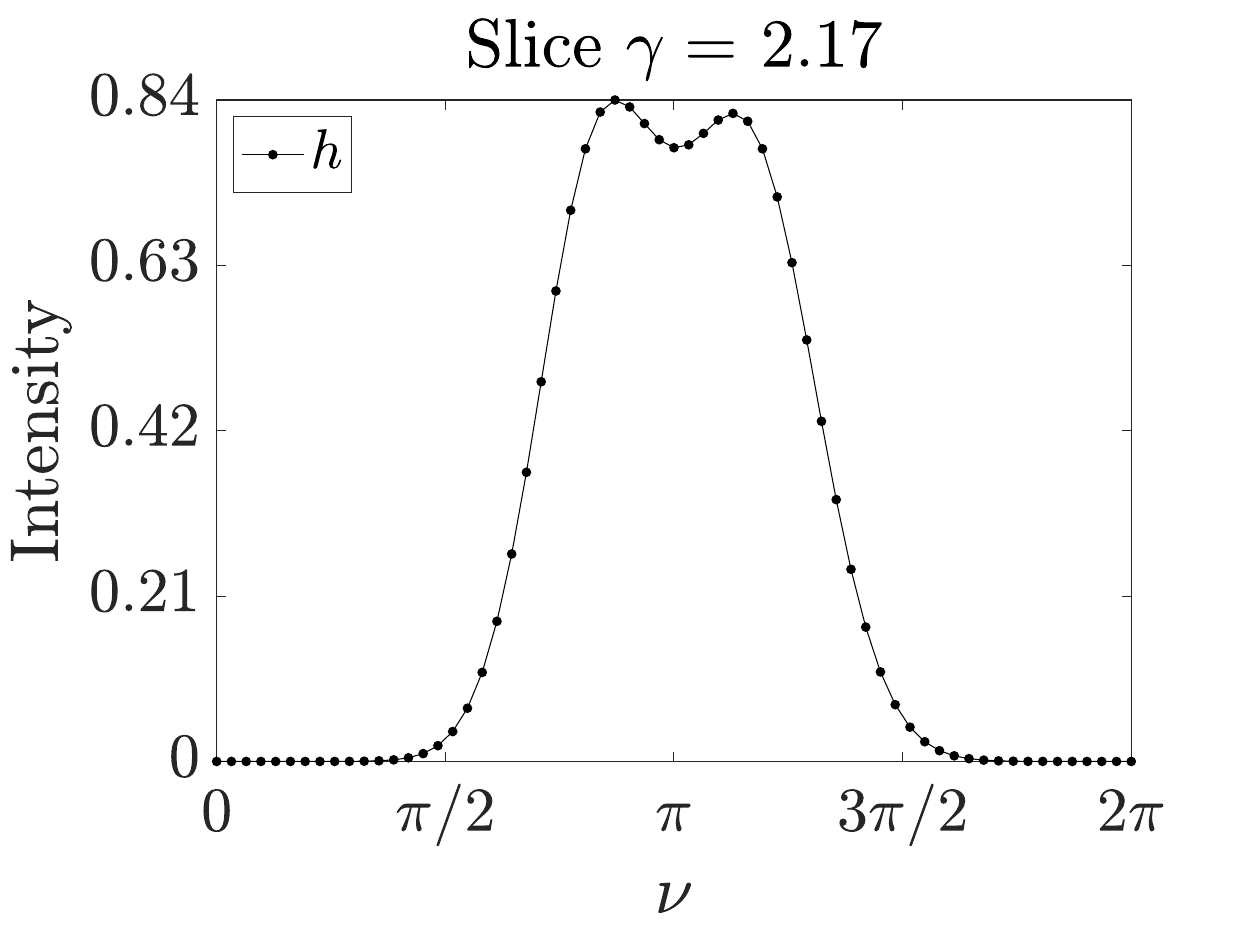}\\[5pt]
	\includegraphics[width=0.27\linewidth]{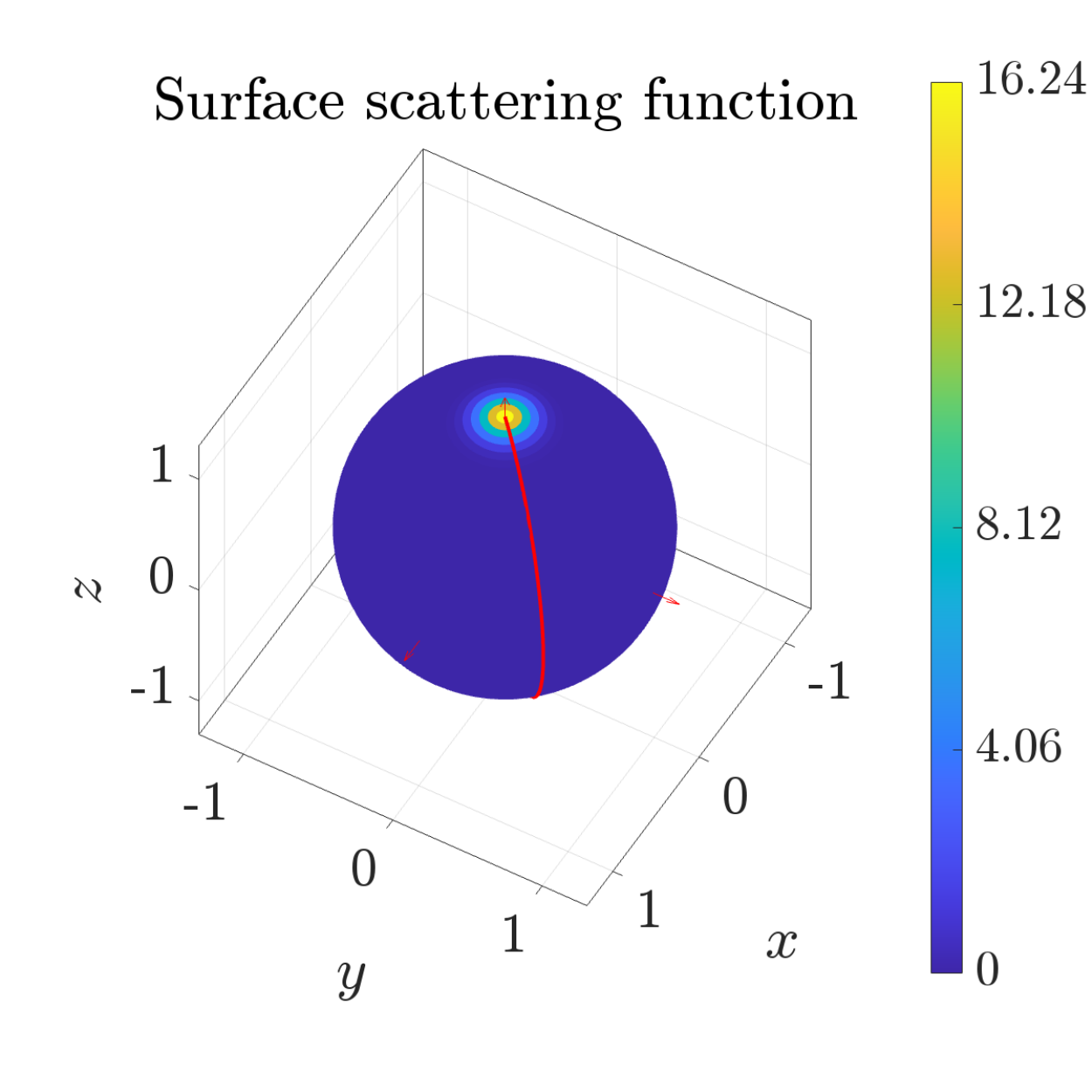}\hspace{10pt}
	\includegraphics[width=0.33\linewidth]{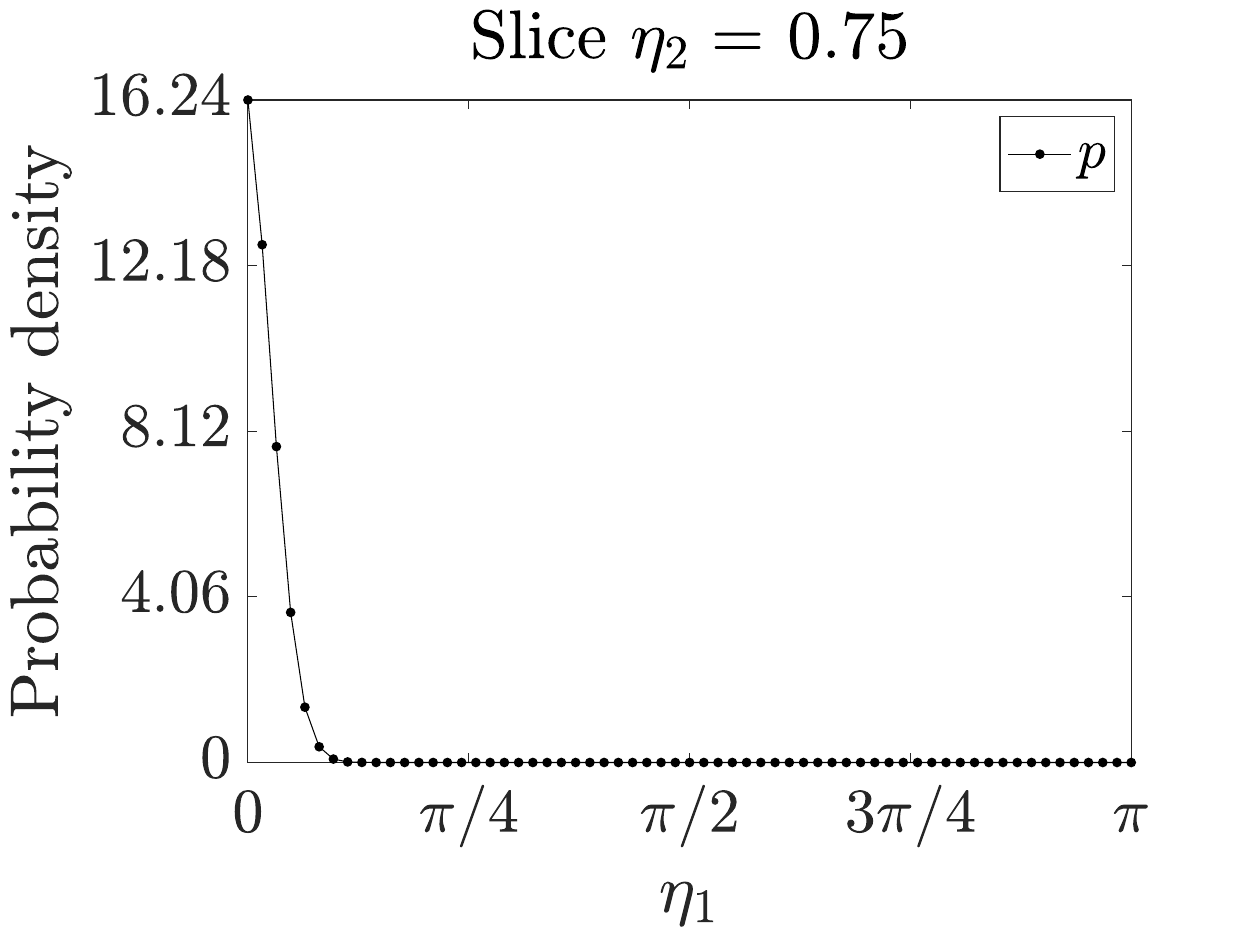}
	\captionsetup{width=\linewidth}
	\caption{Prescribed target distribution and surface scattering function; $64^2$ sample points.}
	\label{fig:example_1-1}
\end{figure*}

This section considers a numerical example to verify our proposed algorithm for computing freeform reflectors with a scattering surface.
The problem is as follows: a uniform parallel source with exitance $f(x,y) = 0.25$ W/m$^2$ on $\mcS = [-1,1] \times [-1,1]$ is to be transformed into three partially overlapping Gaussians in the far field.
The target intensity distribution, $h$, and \textit{surface scattering function}, $p$, are plotted in Fig.~\ref{fig:example_1-1}.

\begin{figure*}[htb!]
	\includegraphics[width=0.33\linewidth]{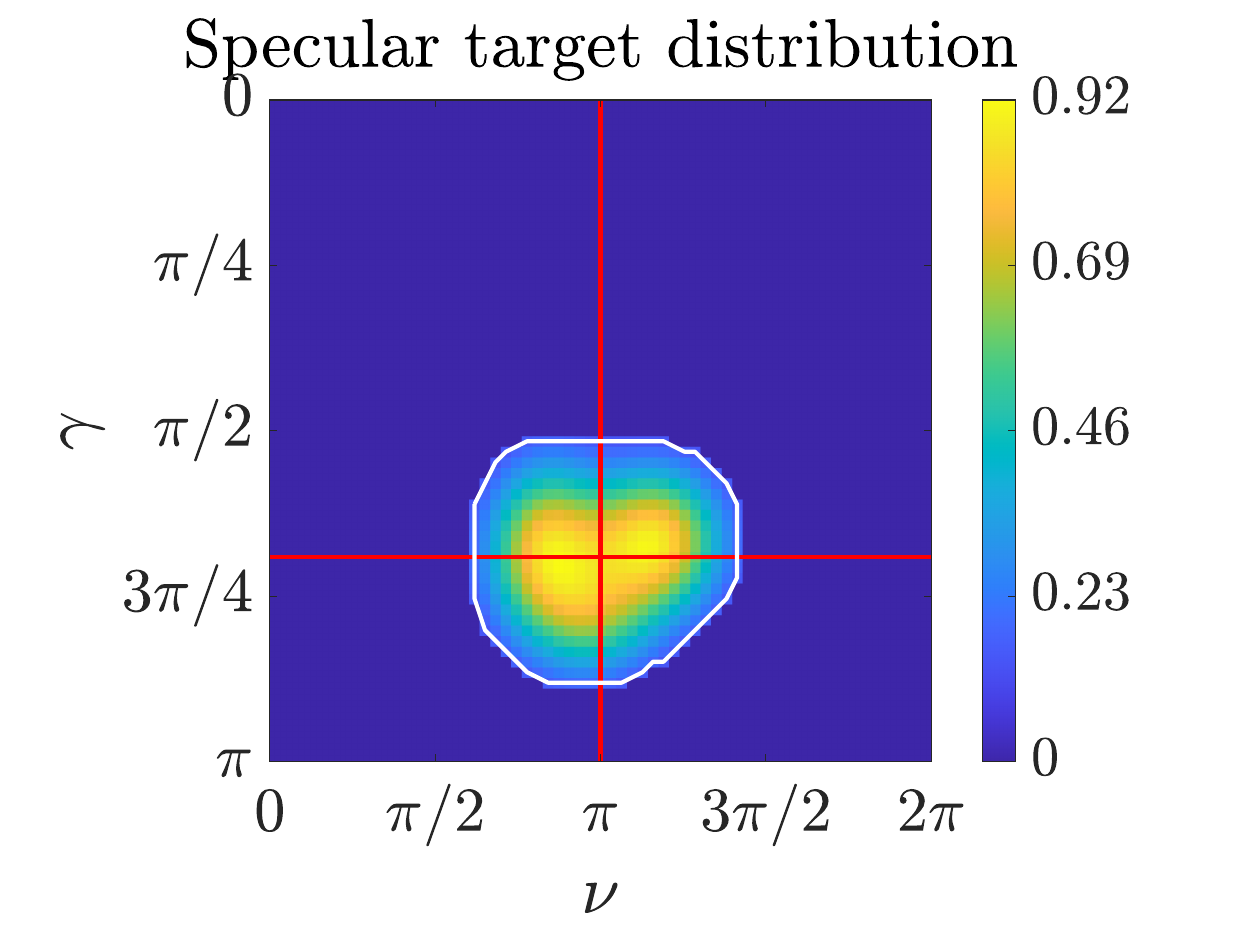}\hfill
	\includegraphics[width=0.33\linewidth]{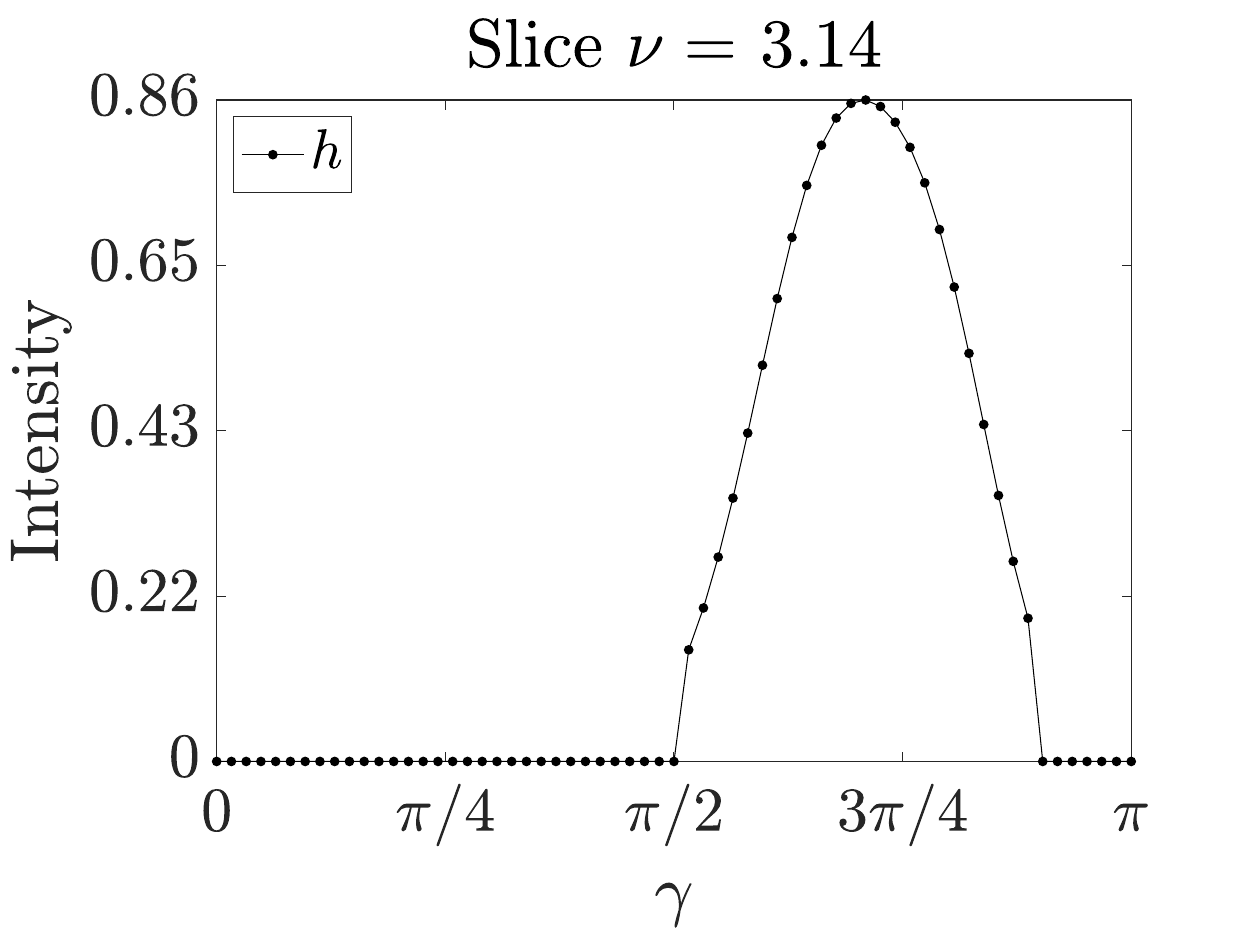}\hfill
	\includegraphics[width=0.33\linewidth]{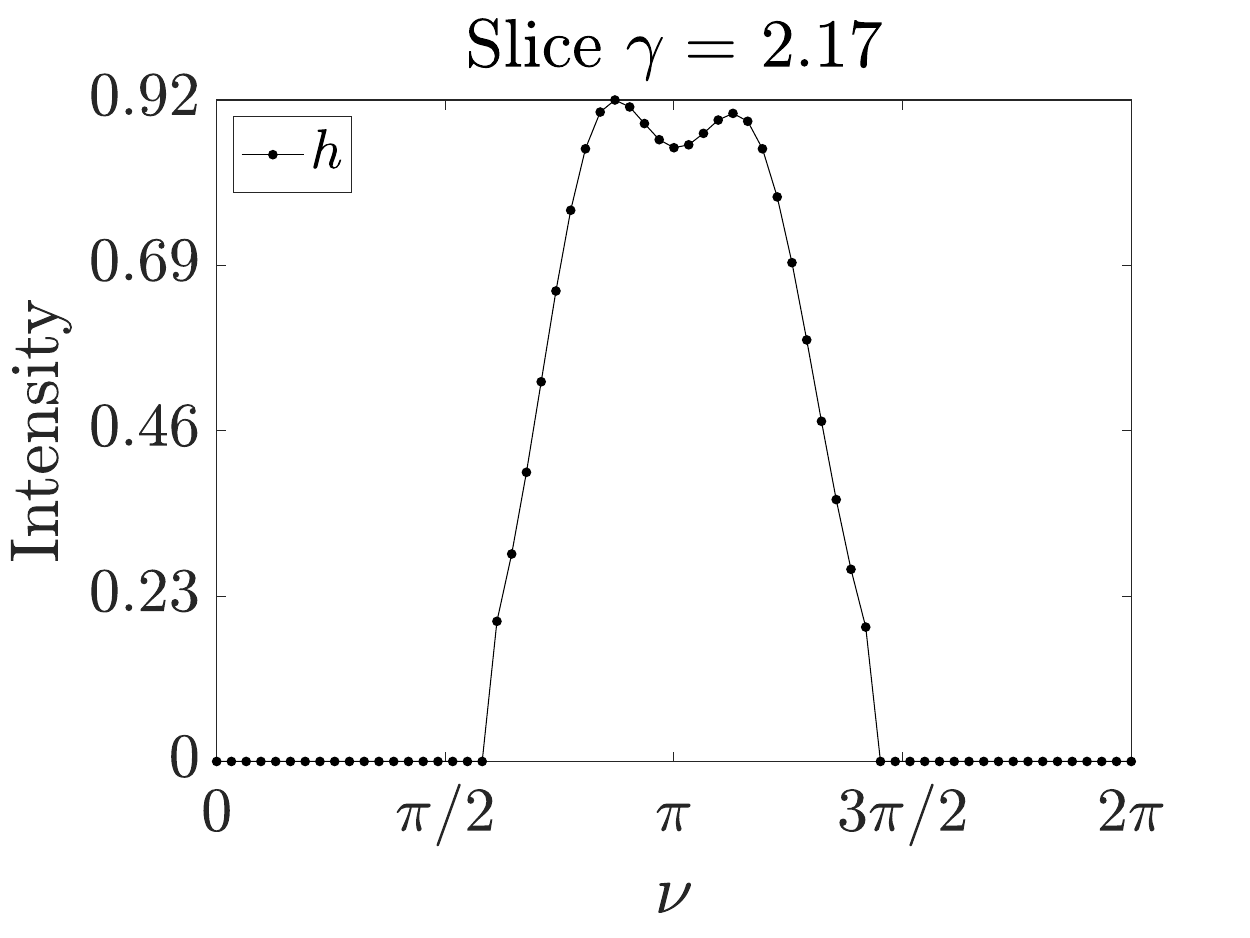}
	\captionsetup{width=\linewidth}
	\caption{The target domain $\mcU$ (white line) and the renormalized target distribution; cutoff value $\epsilon = 0.2\max(h)$.}
	\label{fig:example_1-2}
\end{figure*}

The first step is to compute the \textbf{specular} reflector that transforms $f$ into $h$.
To do so, we must find the target domain $\mcU = \{(\gamma,\nu)\ |\ h(\gamma,\nu) > \epsilon\}$ since Gaussians do not have finite support, which is a requirement for the least-squares solver we used to compute the specular freeform reflector.
We used $\epsilon = 0.2\max(h)$ in this example.
This was achieved using Matlab's \texttt{boundary} routine.
After this step, we renormalized $h$ to unity to guarantee energy conservation since the source has unit flux, i.e.,
\begin{equation}
	\int_{\mcS} f(x,y) \, \dd x \, \dd y = \int_{\mcU} h(\gamma,\nu) \, \sin(\gamma) \, \dd\gamma \, \dd\nu = 1.
\end{equation}
The target domain $\mcU$ is shown as the white curve in Fig.~\ref{fig:example_1-2}.

\begin{figure*}[htb!]
	\centering
	\includegraphics[width=0.33\linewidth]{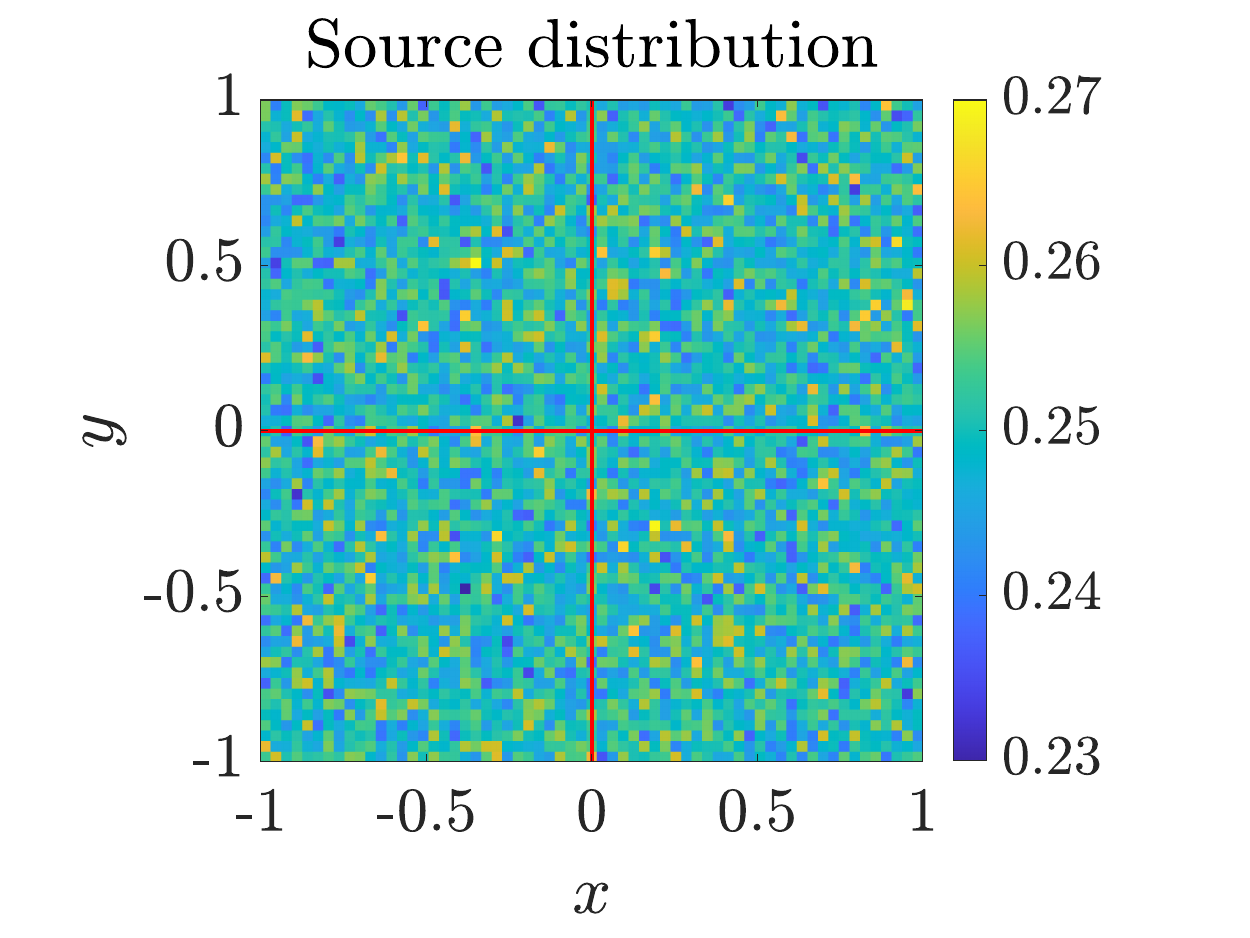}\hfill
	\includegraphics[width=0.33\linewidth]{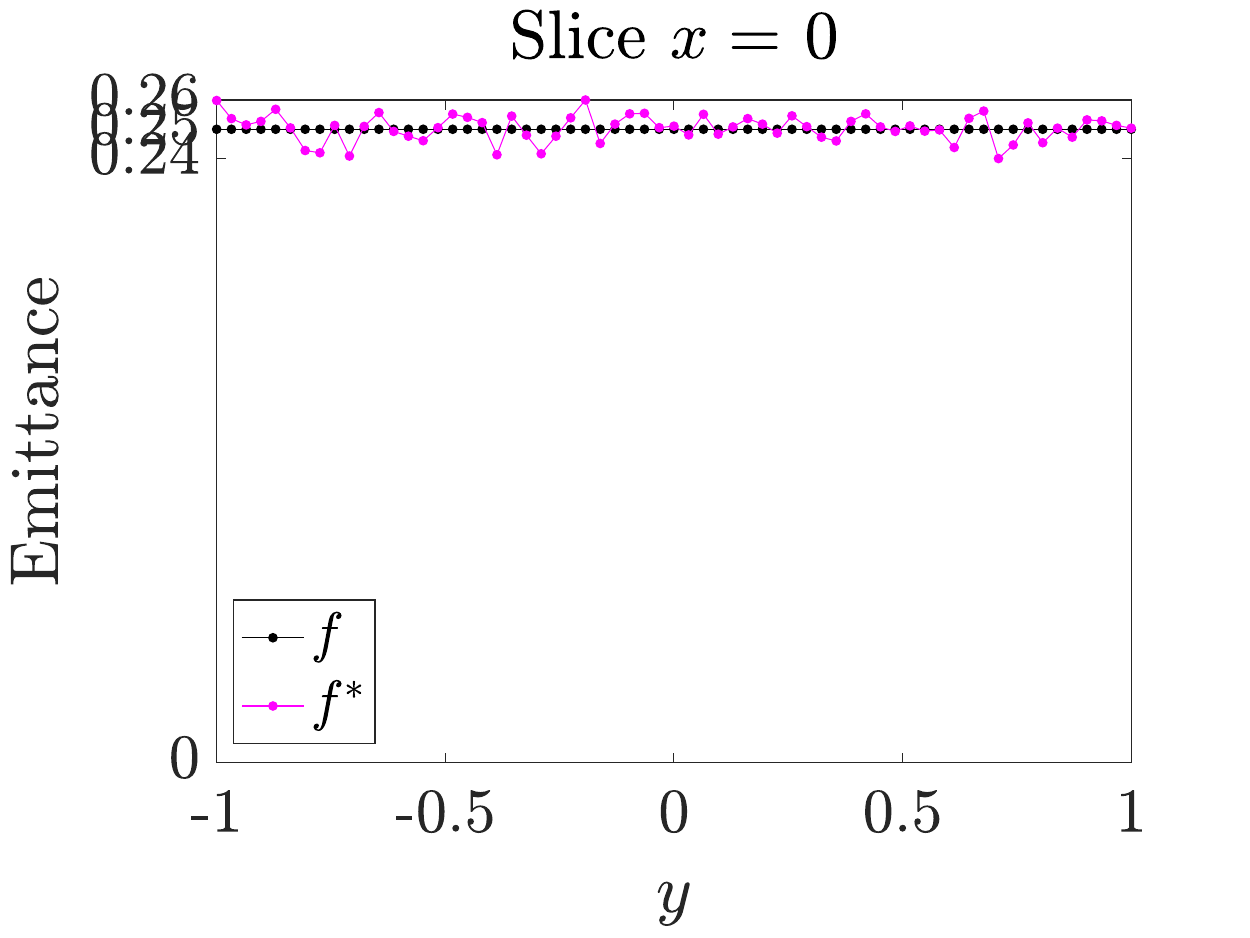}\hfill
	\includegraphics[width=0.33\linewidth]{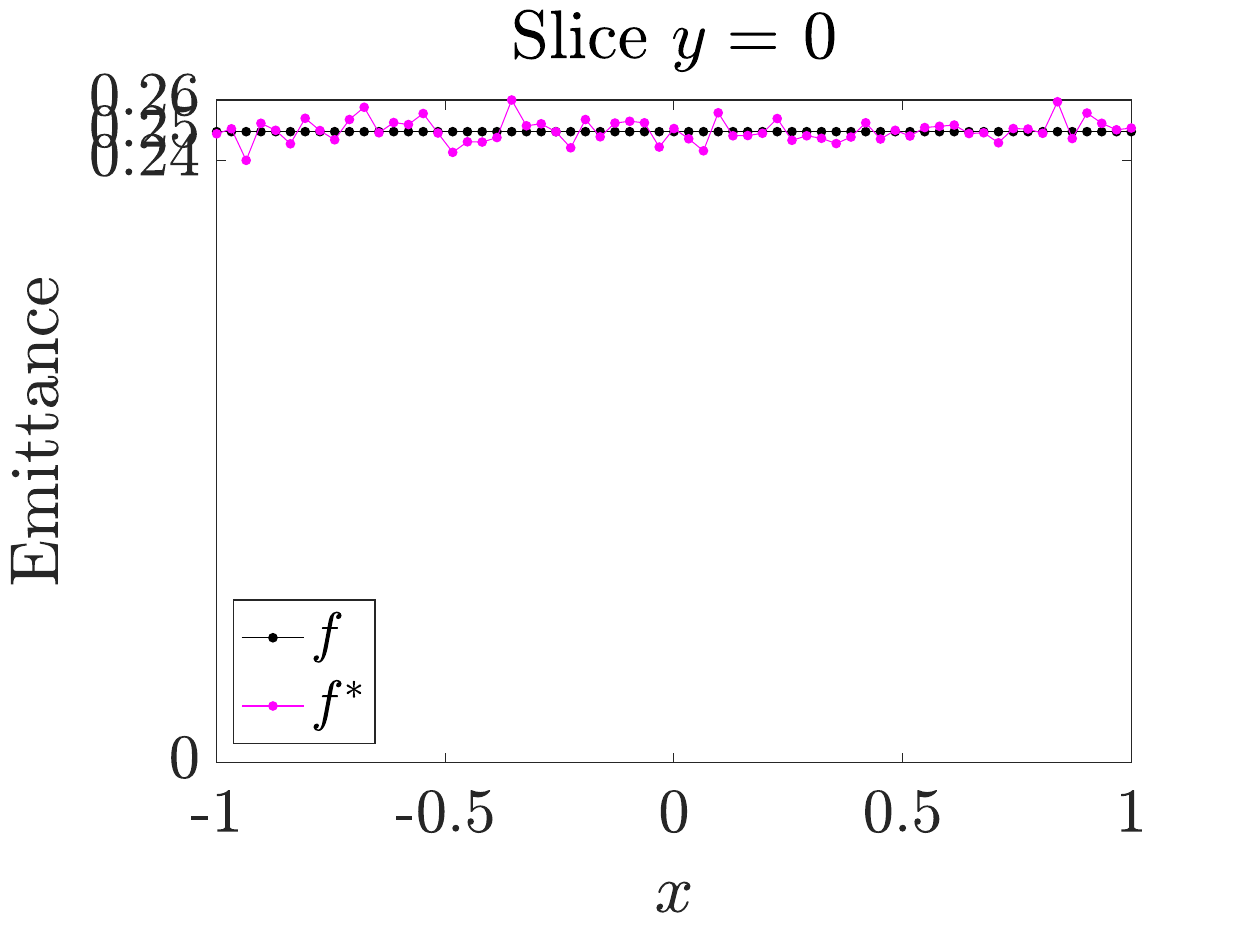}\\[5pt]
	\includegraphics[width=0.33\linewidth]{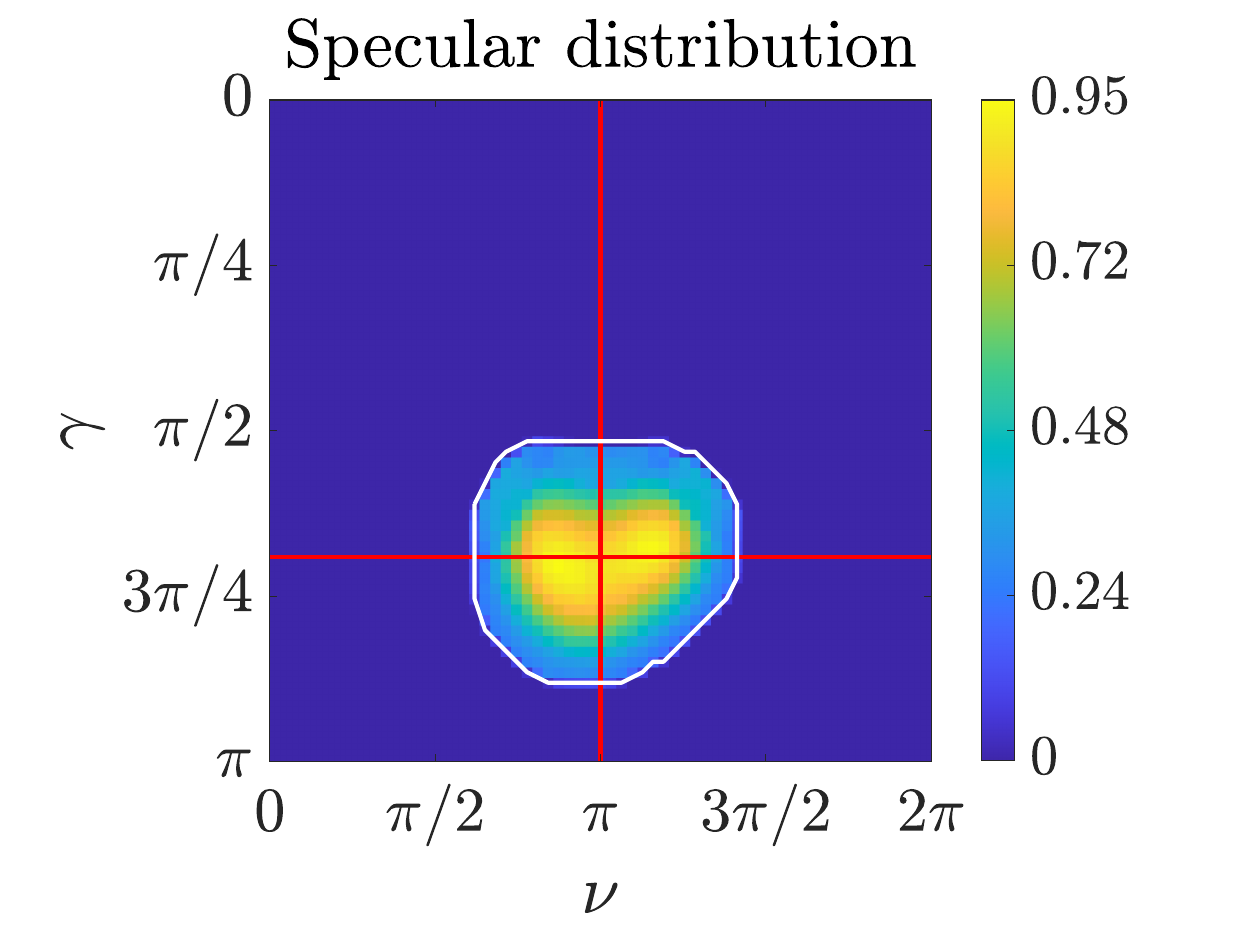}\hfill
	\includegraphics[width=0.33\linewidth]{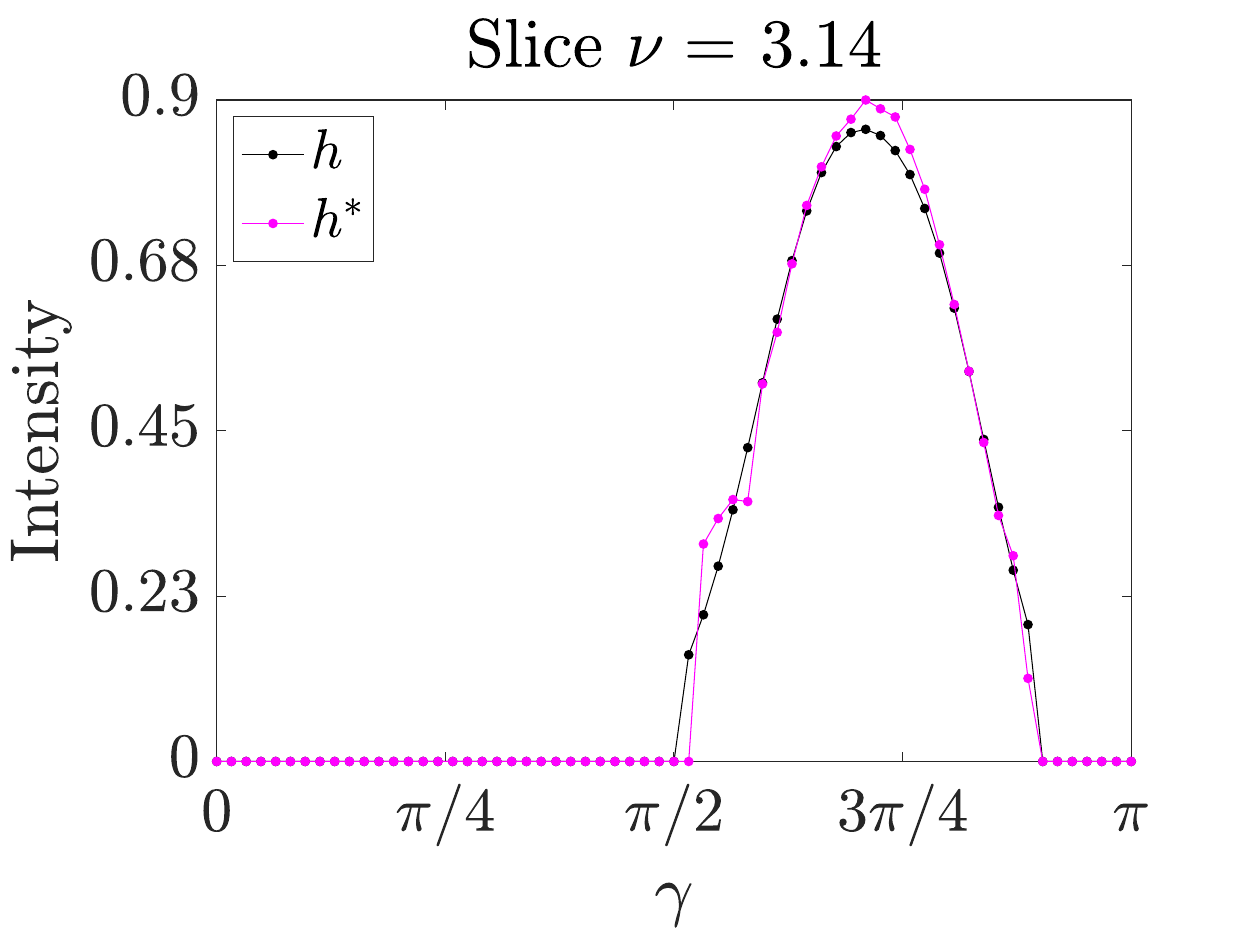}\hfill
	\includegraphics[width=0.33\linewidth]{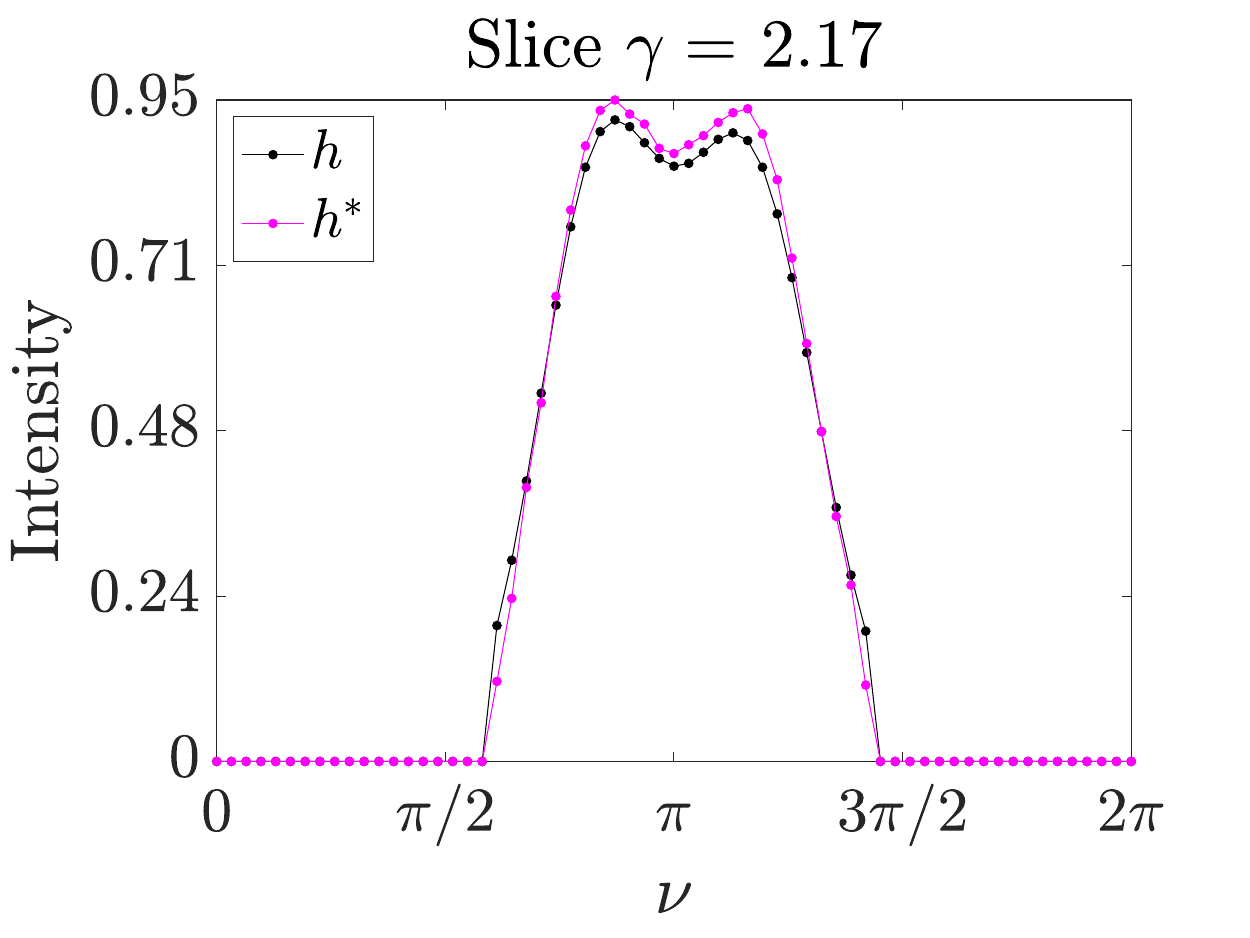}
	\captionsetup{width=\linewidth}
	\caption{Raytraced source and target distributions of the initial reflector; $10^7$ rays traced and $63^2$ collection bins.}
	\label{fig:example_1-5}
\end{figure*}

Next, we computed the specular freeform reflector (not shown), which we then raytraced using our raytracer to see how close the specular distribution was to the renormalized $h$ from Fig.~\ref{fig:example_1-2}.
The resulting distributions are shown in Fig.~\ref{fig:example_1-5}.
Here, the specular reflector performs relatively poorly at recreating $h$.
Still, the specular reflector is not the main star of the show.
Recall that its only use is as a starting point for the minimization procedure.

\begin{figure*}[htb!]
	\centering
	\includegraphics[width=0.33\linewidth]{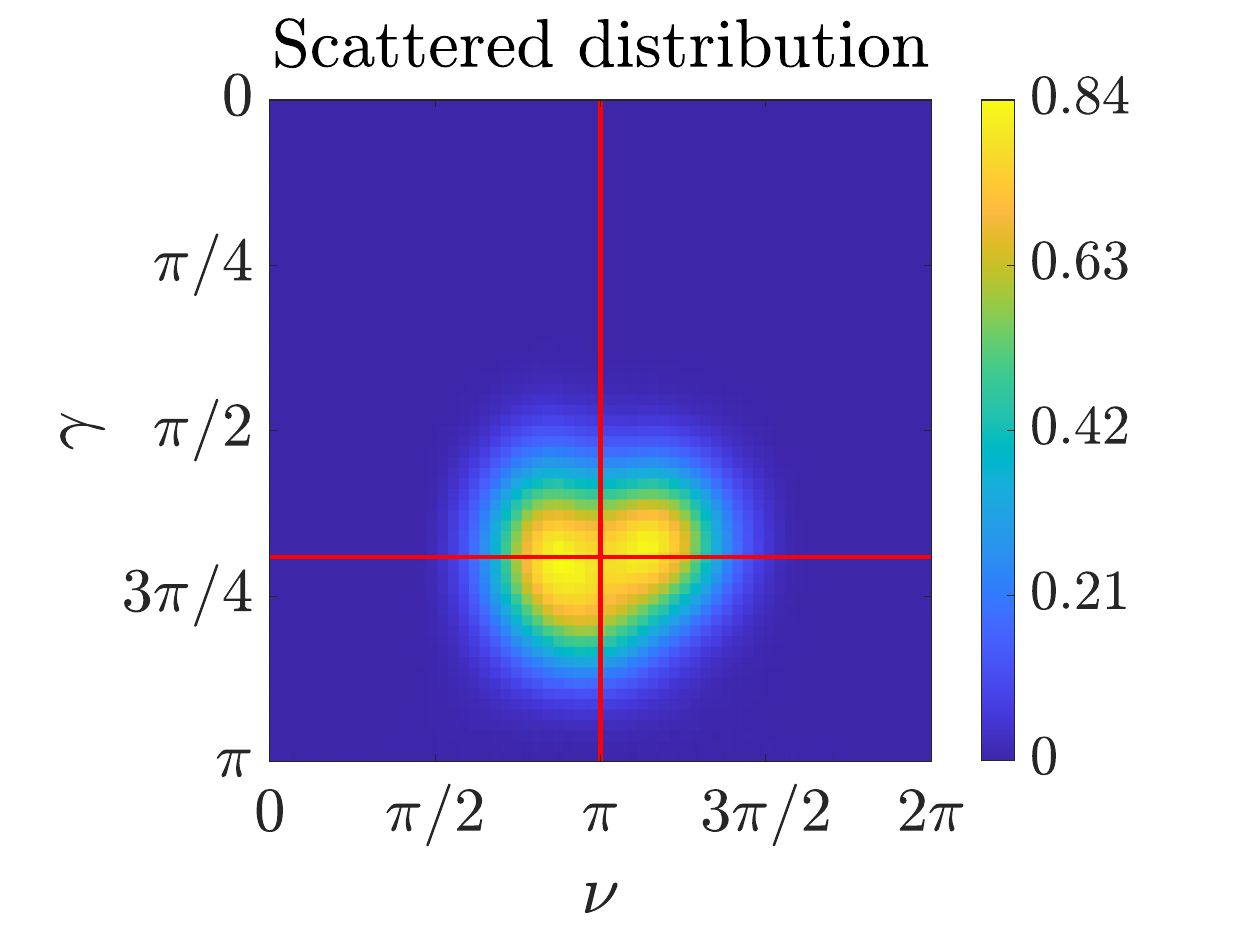}\hfill
	\includegraphics[width=0.33\linewidth]{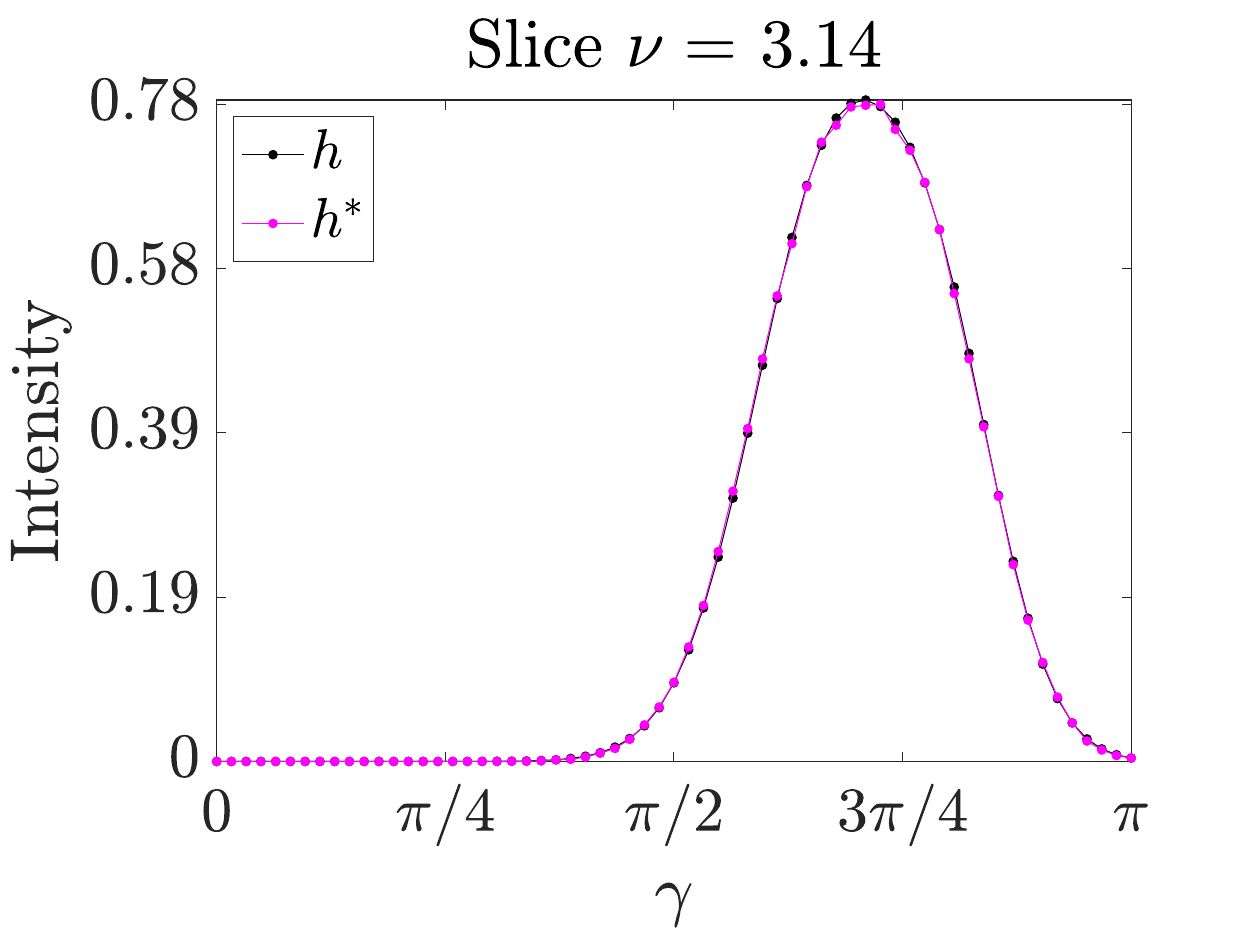}\hfill
	\includegraphics[width=0.33\linewidth]{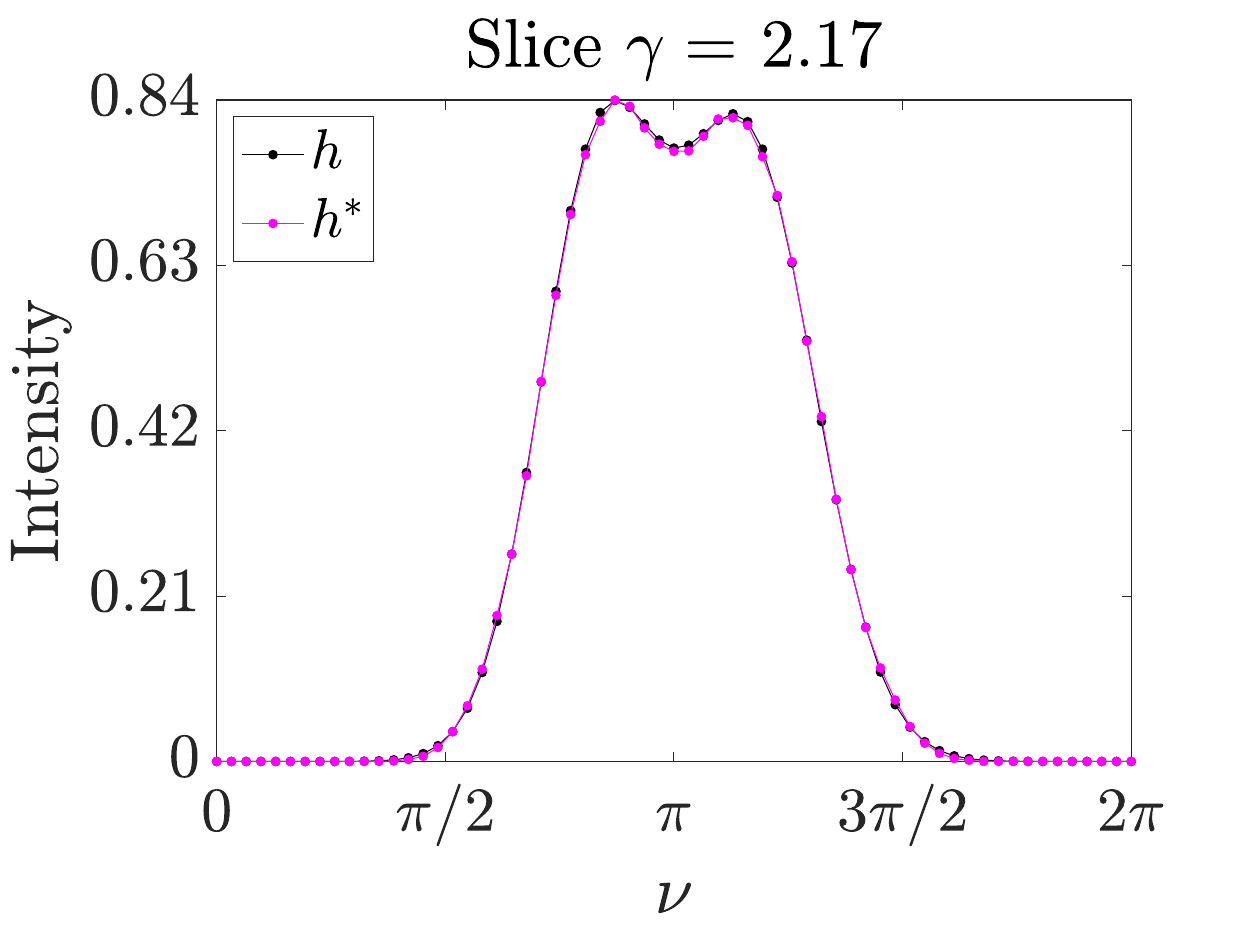}\\[5pt]
	\includegraphics[width=0.33\linewidth]{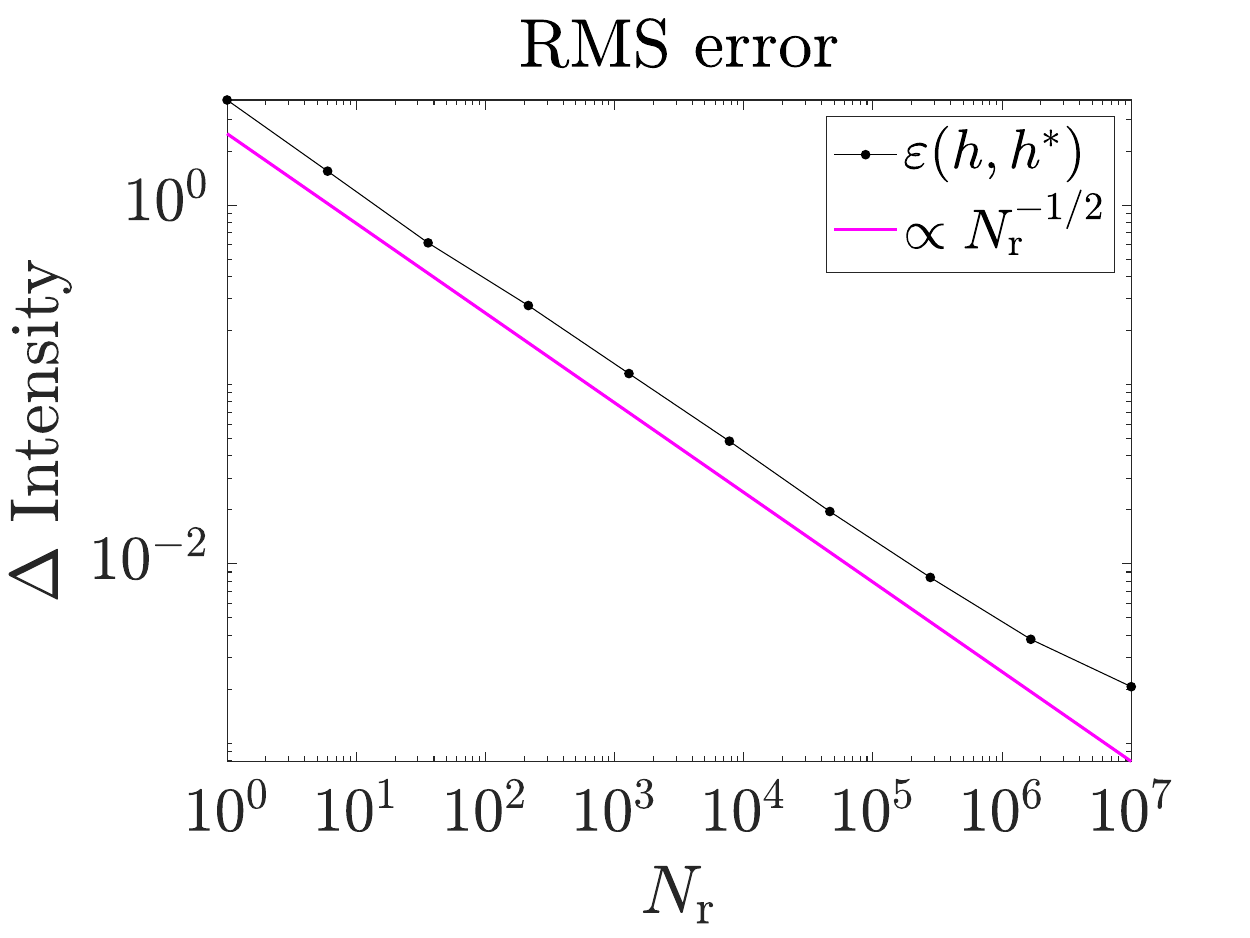}
	\captionsetup{width=\linewidth}
	\caption{Raytraced target distribution after 30 minimization iterations using the optimized normals; $10^7$ rays traced and $63^2$ collection bins.}
	\label{fig:example_1-6}
\end{figure*}

After a mere $30$ iterations of using Matlab's \texttt{fminunc} routine, the optimized normals readily produced the prescribed $h$ when taking scattering into account --- see Fig.~\ref{fig:example_1-6}.
Evidently, the minimization worked very well.
Note in particular the near-perfect $N_{\rmr}^{-1/2}$-convergence, as expected for Monte Carlo raytracing \cite[p.~9]{filosaPhaseSpaceRay2018}.

\begin{figure*}[htb!]
	\centering
	\includegraphics[width=0.3\linewidth]{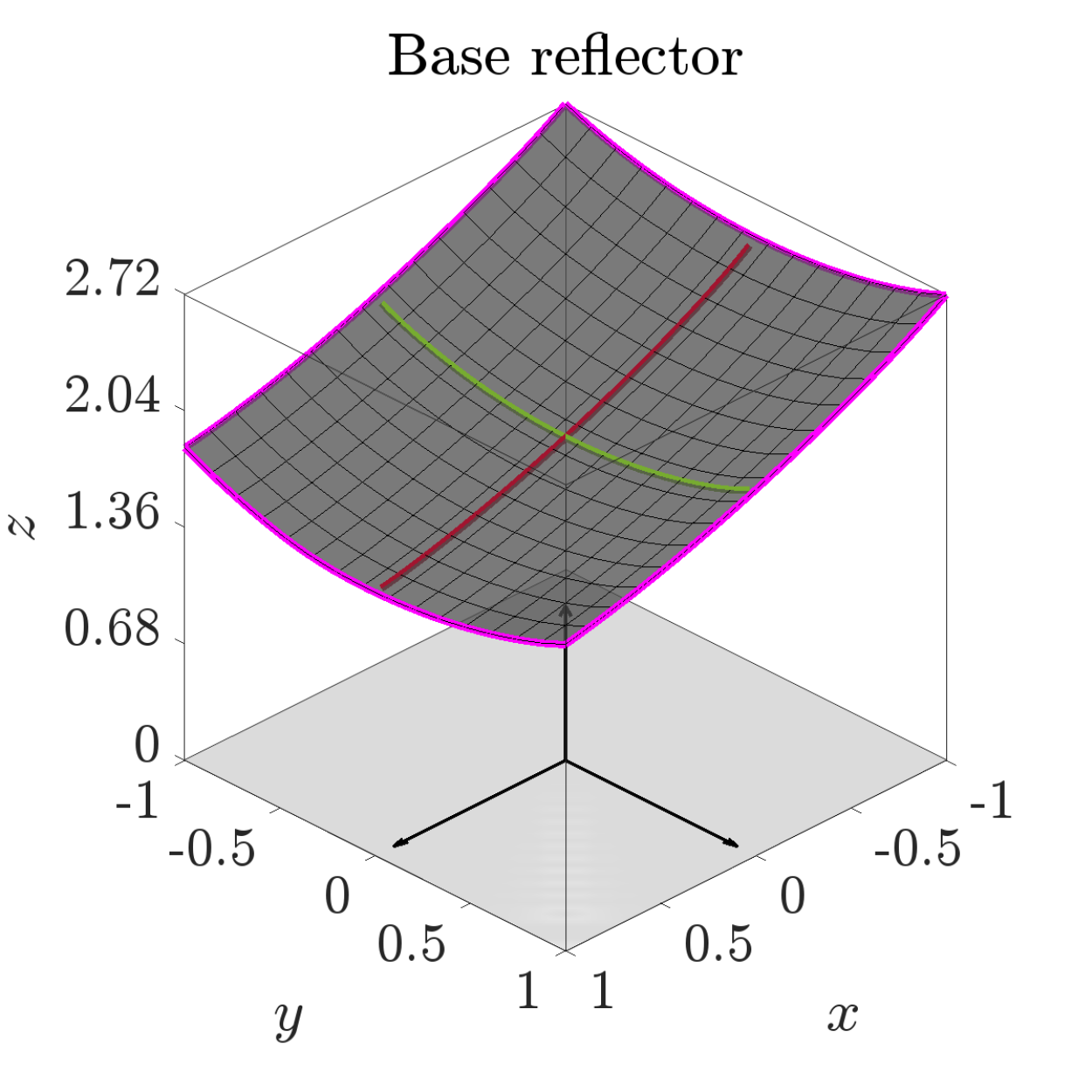}\hfill
	\includegraphics[width=0.33\linewidth]{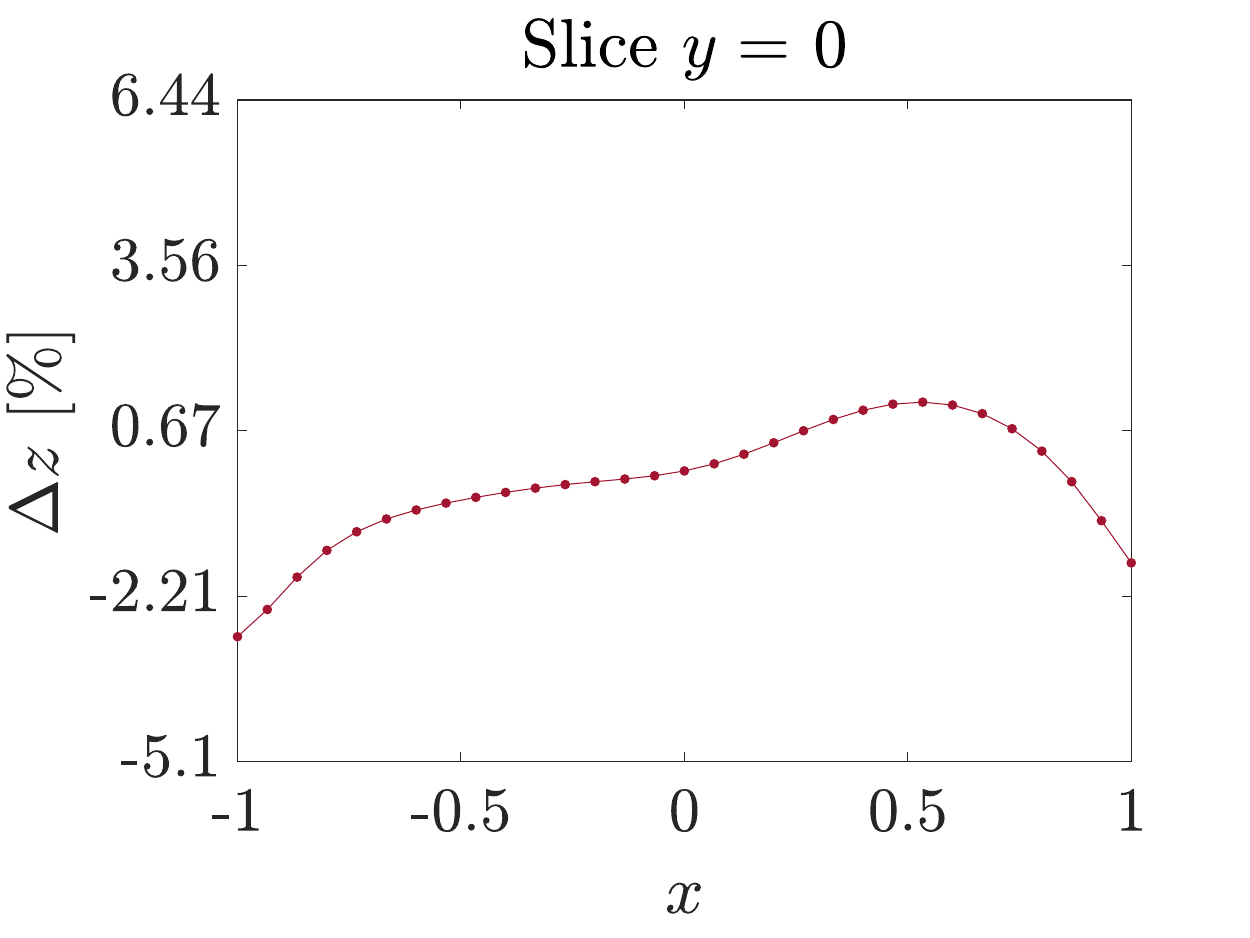}\hfill
	\includegraphics[width=0.33\linewidth]{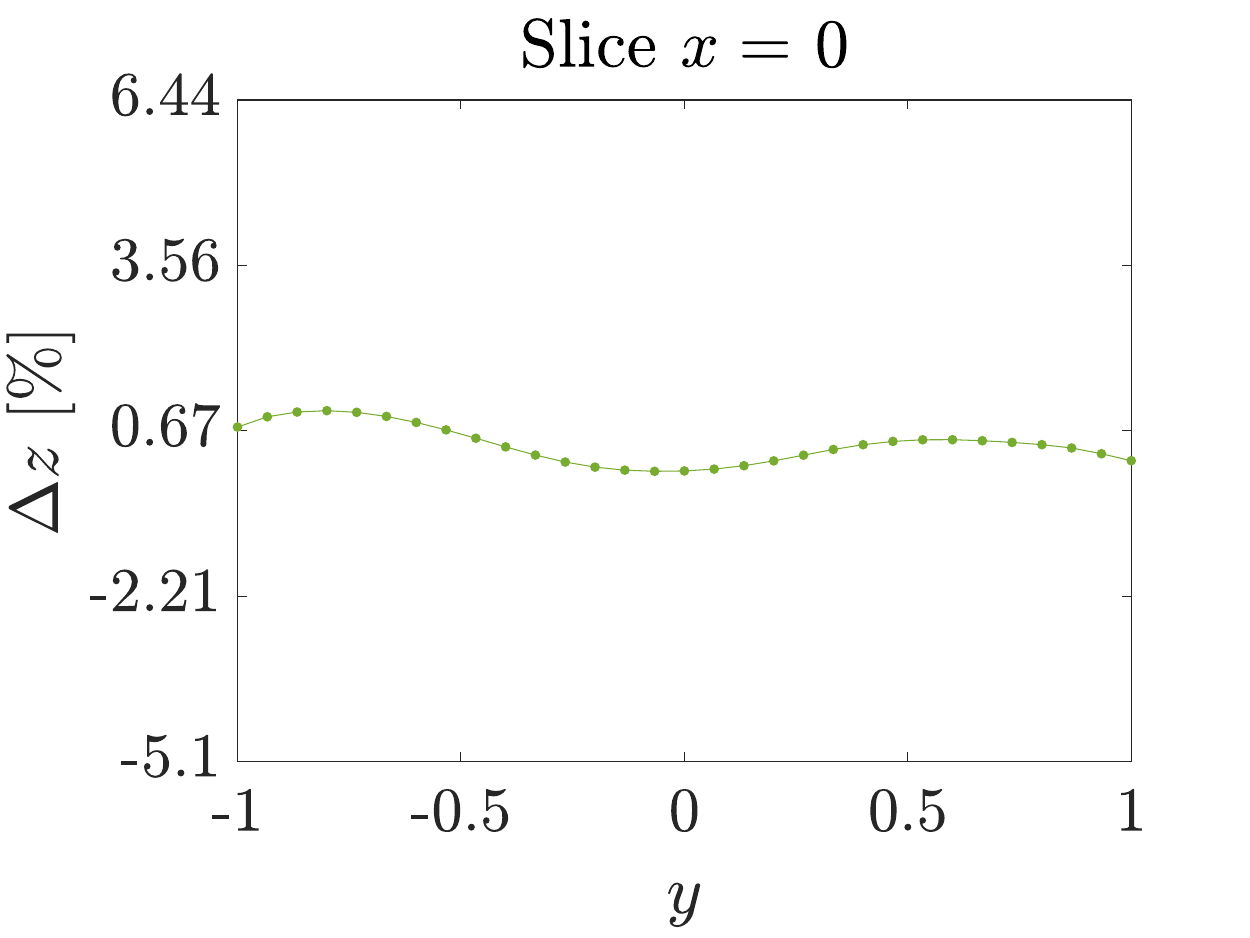}\\[5pt]
	\includegraphics[width=0.3\linewidth]{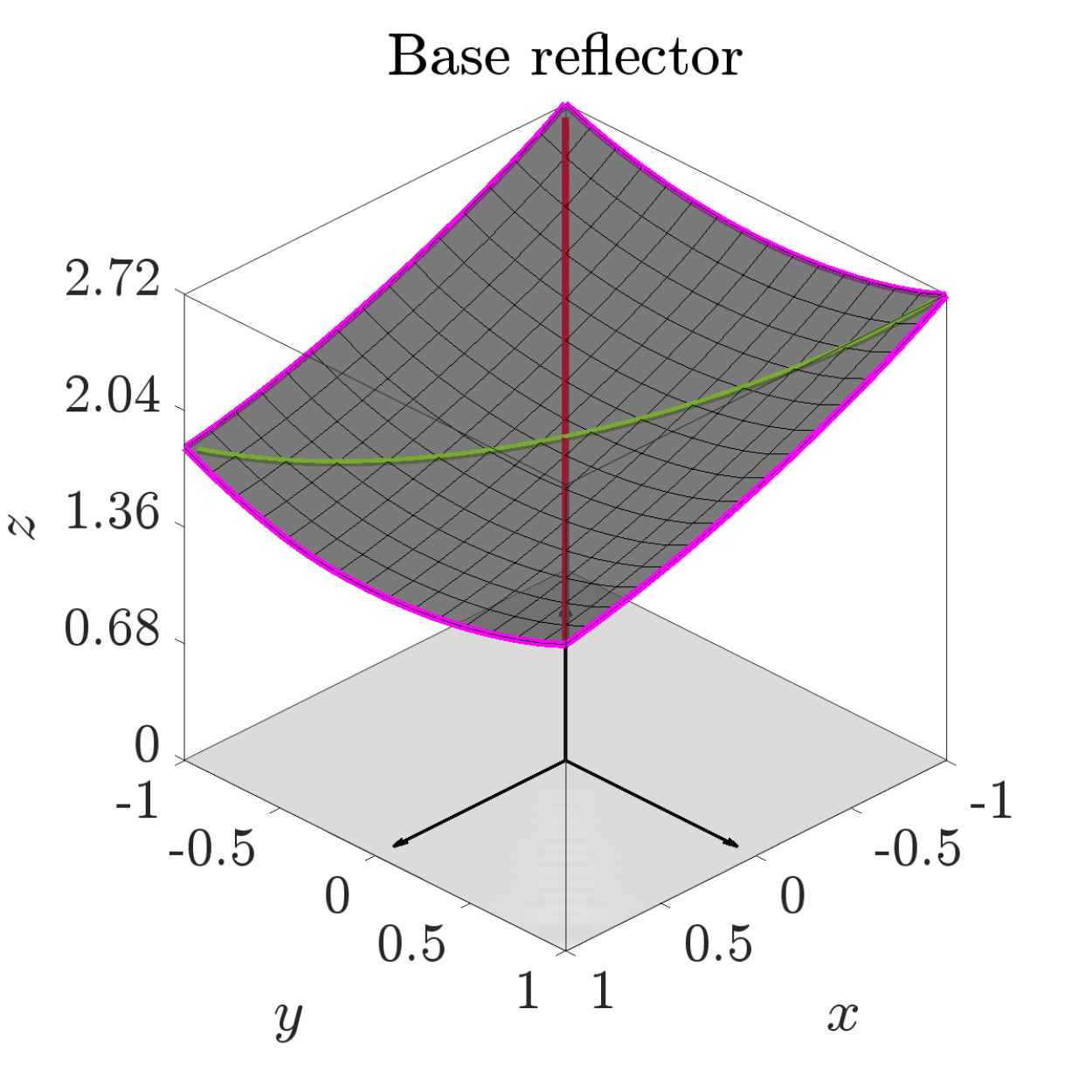}\hfill
	\includegraphics[width=0.33\linewidth]{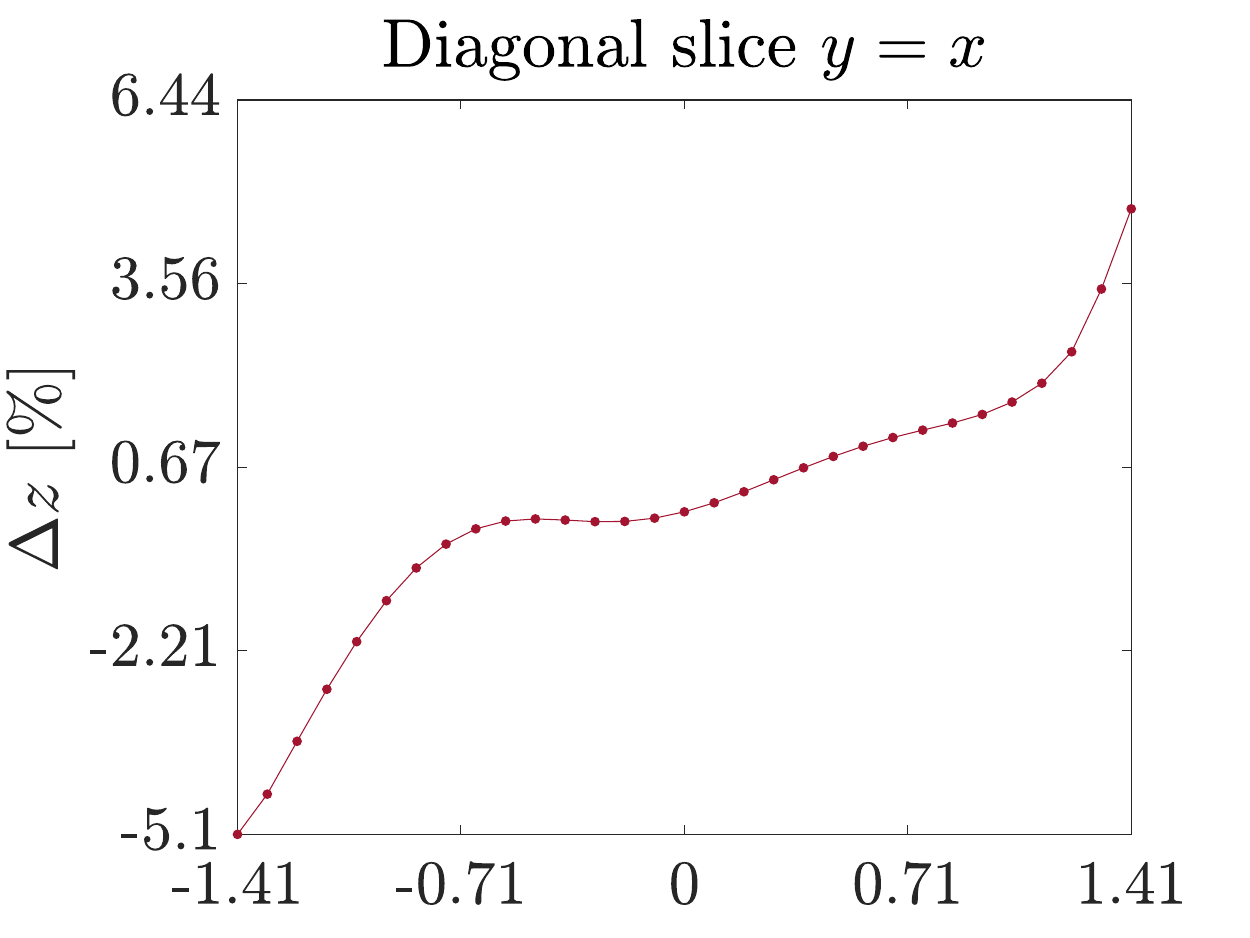}\hfill
	\includegraphics[width=0.33\linewidth]{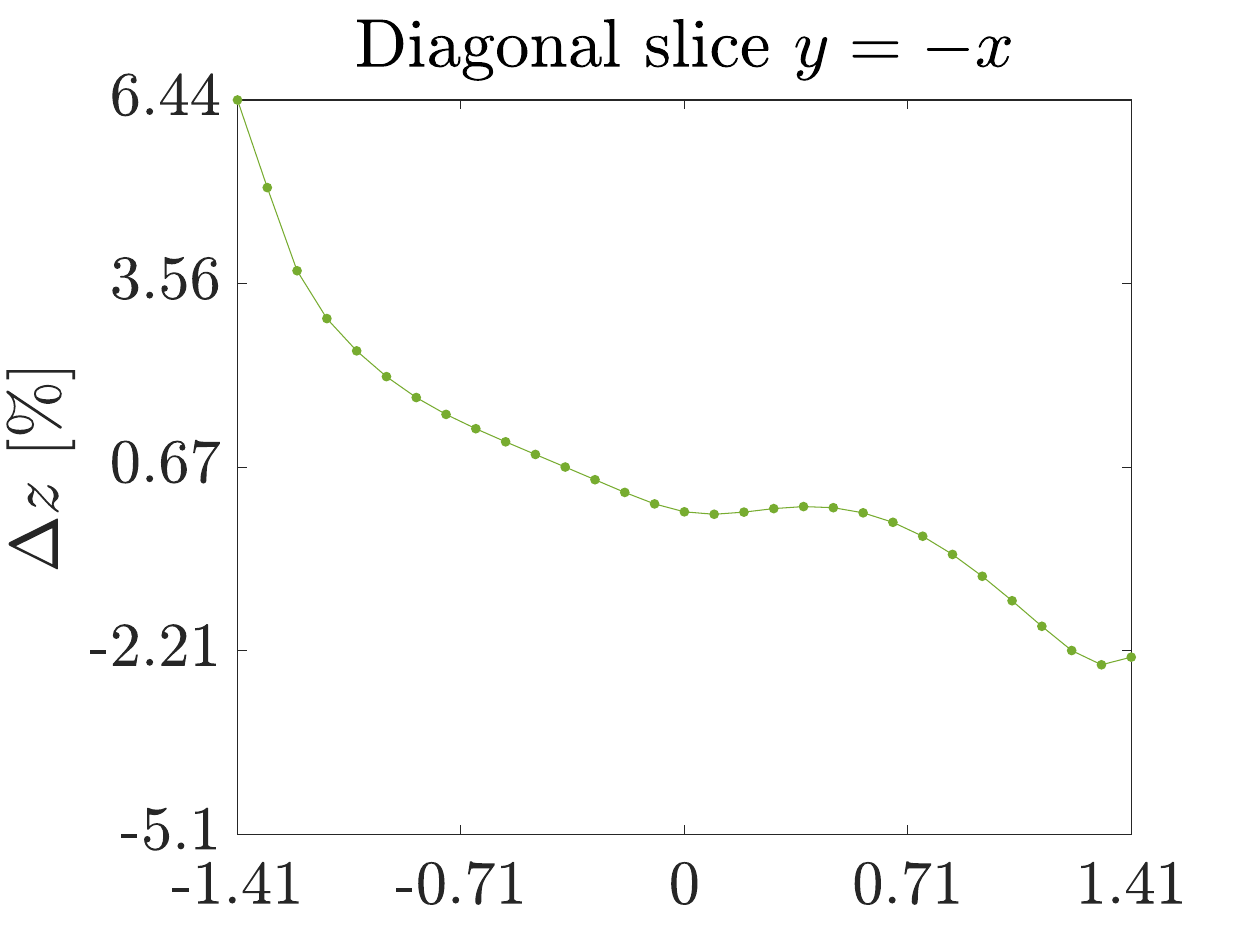}
	\captionsetup{width=\linewidth}
	\caption{Slices along the indicated lines of the difference in reflector height before and after optimizing the normals; $32^2$ sample points.}
	\label{fig:example_1-7}
\end{figure*}

Let us now investigate the changes in the reflector surface due to the minimization procedure.
First, let
\begin{equation}\label{eq:deltaZ}
	\Delta z := \frac{z - z_\rmb}{u_0} \cdot 100,
\end{equation}
where $z_\rmb$ and $z$ are the heights of the specular (\textit{base}) reflector and the reflector with nonzero scattering, respectively, and $u_0 = 1$ represents the average offset such that $\Delta z$ becomes a percentage.
The result is shown in Fig.~\ref{fig:example_1-7}, where we see relative differences in reflector height of a few percent, typically considered manufacturable in illumination optics.

\clearpage
\section{Conclusions \& Discussion}\label{sec:conclusions}
We have developed a simple model of surface roughness based on microfacet surface roughness and shown how to use it to compute freeform reflectors with a scattering surface in the context of illumination optics.
Starting with a specular reflector, we alter it using a minimization procedure to account for scattering such that the new reflector results in the desired target intensity when taking scattering into account.
The problem thus reduces to computing a specular reflector, which can be done using methods from literature and then altering it using a typical minimization procedure.
We demonstrated the proposed solution algorithm with a concrete numerical example where we computed a freeform reflector surface that converted a parallel bundle of rays into three partially overlapping normal distributions.

As we hinted towards when developing the microfacet-based scattering model, this approach is favorable compared to our previous work in \cite{kronbergThreedimensionalFreeformReflector2023}.
Specifically, it seems more straightforward to generalize the solution algorithm to non-isotropic surfaces, starting from Eq.~\eqref{eq:hGamma_Li}.
Additionally, more complicated BRDFs can be substituted into our expressions to improve our model.\\

\hrule
\begin{changemargin}{1cm}{1cm}
	\noindent\textsc{\textbf{Funding:}} This work was partially supported by the Dutch Research Council (\textit{Dutch:} Nederlandse Organisatie voor Wetenschappelijk Onderzoek (NWO)) through grant P15-36.\\
	\textsc{\textbf{Disclosures:}} The authors declare no conflicts of interest.\\
	\textsc{\textbf{Data availability:}} Data underlying the results presented in this paper are not publicly available at this time but may be obtained from the authors upon request.
\end{changemargin}
\hrule

\pagestyle{ref}
\addcontentsline{toc}{section}{References}
\printbibliography

\end{document}